\documentclass[aps,prd,preprintnumbers,reprint,nofootinbib,showpacs,onecolumn,notitlepage]{revtex4-1}
\usepackage{graphicx,bm,amssymb,amsmath}
\usepackage{color}

\newcommand{\defeq}{\mathrel{\mathop:}=}

\begin{document}

\title{\boldmath Electroweak Sudakov effects in $W$, $Z$ and $\gamma$ production at large
transverse momentum}
\author{Thomas Becher}
\affiliation{Albert Einstein Center for Fundamental Physics, Institut f\"ur Theoretische Physik, Universit\"at Bern,
  Sidlerstrasse 5, CH-3012 Bern, Switzerland}
\author{Xavier \surname{Garcia i Tormo}\\ \phantom{a}}
\affiliation{Albert Einstein Center for Fundamental Physics, Institut f\"ur Theoretische Physik, Universit\"at Bern,
  Sidlerstrasse 5, CH-3012 Bern, Switzerland}

\date{\today}


\begin{abstract}
We study electroweak Sudakov effects in single $W$, $Z$ and $\gamma$ production
at large transverse momentum using Soft Collinear
Effective Theory. We present a factorized form of the cross section
near the partonic threshold with both QCD and electroweak effects included and compute the electroweak corrections arising at different scales. We analyze their size relative to the QCD corrections
as well as the impact of strong-electroweak mixing terms.  Numerical results for the vector-boson cross sections at the Large Hadron Collider are presented.
\end{abstract}


\maketitle

\section{Introduction}\label{sec:intro}
The production of a single electroweak boson is one of the basic
hard-scattering processes that one can measure at hadron
colliders. Much theoretical effort has been put over the years into
precisely predicting the cross section of the $W$, $Z$,
and $\gamma$ production processes. There is ongoing work to obtain the second-order corrections in the strong coupling $\alpha_s$ to the transverse momentum spectrum, a quantity for which the next-to-leading order (NLO) results are known for a long time \cite{Ellis:1981hk,Arnold:1988dp,Gonsalves:1989ar} and have been implemented in numerical integration programs \cite{QT,MCFM,Melnikov:2006kv,Catani:2009sm,Gavin:2010az}. One way to improve the fixed-order results is to include resummation of
higher-order terms that are enhanced in certain kinematical limits. Here, we focus on the region of large transverse momentum $p_T$ and compute the cross section near the partonic threshold. In this region, two types of Sudakov-enhanced terms arise, whose combined resummation is the subject of the present paper. First of all, the electroweak corrections are enhanced by double logarithms of the vector-boson masses $M_W$ and $M_Z$ over $p_T$. Secondly,  near threshold, the invariant mass of the hadronic jet which recoils against the electroweak boson is small and the perturbative corrections are enhanced by logarithms of the jet mass $M_X$ over $p_T$. 
One can expand around the threshold limit and resum the enhanced
terms. For the electroweak-boson spectrum, threshold resummation was first achieved at next-to-leading logarithmic (NLL) accuracy in \cite{Laenen:1998qw}. Except in the unrealistic case where the $p_T$ value is close to the maximum kinematically-allowed value, the cross section also receives contributions away from the threshold region. However, the partonic threshold contributions often amount
to the bulk of the hadronic cross section. This is due to the rapid fall-off of the parton distribution functions (PDFs) $f(x)$ at large $x$, which dynamically enhances the threshold region (see 
Ref.~\cite{Becher:2007ty} and references therein).

In the partonic threshold limit, the real radiation simplifies
considerably because of the restricted phase space. The hadronic final
state consists of the electroweak vector boson recoiling against a single low-mass
jet, and all additional hadronic radiation must be either soft, or
collinear to the jet or the incoming hadrons. This kinematical
situation is amenable to an effective theory treatment using Soft Collinear Effective Theory (SCET)
\cite{Bauer:2000yr,Bauer:2001yt,Beneke:2002ph}. Within the SCET
framework, threshold resummation of quantum chromodynamics (QCD)
corrections for $W$, $Z$ and $\gamma$ production at large $p_T$ has been
achieved at next-to-next-to leading logarithmic (N$^2$LL) accuracy
\cite{Becher:2009th,Becher:2011fc,Becher:2012xr}. Some results with
N$^2$LL accuracy were also presented in Ref.~\cite{Kidonakis:2012sy},
using the traditional diagrammatic approach to resummation. Essentially all
the ingredients required to achieve next-to-next-to-next-to leading logarithmic
(N$^3$LL) accuracy in the SCET framework are by now known
\cite{Garland:2001tf,Garland:2002ak,Gehrmann:2002zr,Gehrmann:2011ab,Becher:2006qw,Becher:2010pd,Becher:2012za} and
a complete analysis of resummation at N$^3$LL accuracy will be the subject of a future publication. 

At the energies and
luminosities that the Large Hadron Collider (LHC) can reach, also
virtual corrections due to electroweak-boson exchanges can
become quite significant. Since we are considering single
electroweak-boson production, without additional radiation of soft or
collinear $W$ or $Z$ bosons, the cross section will contain logarithms
of the form $\ln(p_T^2/M_{V}^2)$, where $M_V$ is the $W$- or $Z$-boson
mass. This was recognized long ago,
and the electroweak one-loop corrections and two-loop logarithmically
enhanced terms have been computed for these processes
\cite{Kuhn:2004em,Kuhn:2005az,Kuhn:2005gv,Kuhn:2007qc,Hollik:2007sq,Kuhn:2007cv}. The outcome
of these analyses is that electroweak corrections can be as large as
$20\%$ for $p_T\sim 1$~TeV at the LHC, clearly indicating that
electroweak Sudakov effects have to be included if one wants to have a
precise prediction for the spectrum in the region $p_T\gg M_V$. Let us note
that these logarithms would partly cancel if one considered
real $W$ and $Z$ emission, in addition to virtual
electroweak-boson exchanges, but the
cancellation would not be complete, since the initial states
carry non-abelian charge~\cite{Ciafaloni:2000df}. Recently, this was
explicitly verified for the $Z+1$ jet production process at the
double-logarithmic level~\cite{Stirling:2012ak}. In this paper we
elaborate on the inclusion of electroweak effects in the cross sections
using SCET.

\begin{figure}
\centering
\includegraphics[width=10cm]{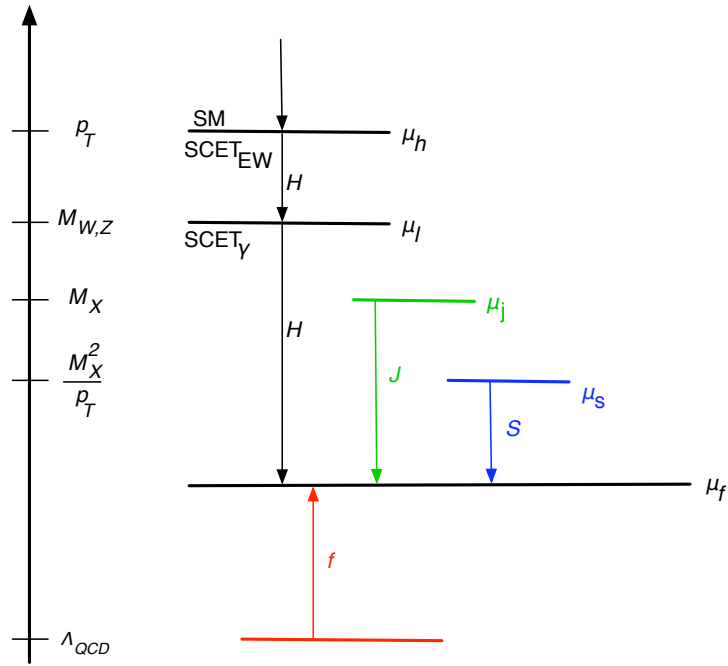}
\caption{Effective theory setup. $H$, $J$, $S$, and $f$ denote the
  hard,  jet, and soft functions, and the PDFs, respectively;
  $\Lambda_{\rm QCD}$ is the QCD scale.}\label{fig:stepsRGev}
\end{figure}

A derivation of the factorization formula for single electroweak-boson
production within SCET, has
been given in Ref.~\cite{Becher:2009th}. The factorization formula will be made more
explicit in the following sections, but schematically we have that
the partonic cross section $\hat{\sigma}$ is given in a factorized
form as 
\begin{equation}\label{eq:schfact}
d \hat{\sigma}\sim \hat{\sigma}^BH\times J_V\otimes J\otimes S,
\end{equation}
where $\hat{\sigma}^B$ is the Born cross section, $H$ the hard function,
and $J$ and $S$ the jet and soft functions, which encode collinear and
soft radiation, respectively. The symbol
$\otimes$ denotes a convolution. Since we will
deal with electroweak corrections, in addition to strong-interaction
effects, we have included a jet function $J_V$ for the electroweak
boson $V=W,Z,\gamma$ in the factorization formula. The formalism
to incorporate electroweak corrections in the SCET framework was
developed in a series of papers by Chiu et al.~\cite{Chiu:2007yn,Chiu:2007dg,Chiu:2008vv,Chiu:2009mg,Chiu:2009ft}. In
those papers, resummation of electroweak Sudakov corrections to the
hard function $H$ was studied in detail, and explicit expressions for
several different hard-scattering processes were given. The strategy to
incorporate electroweak corrections consists of four steps
\cite{Chiu:2009mg}: (i) a matching from the full Standard
Model (SM) to SCET at a high scale $\mu_h\sim p_T$. (ii) Running from $\mu_h$ to a
low scale $\mu_l\sim M_V$. (iii) Matching at the scale $\mu_l$, from a
version of SCET that contains dynamical $Z$ and $W$ bosons to a version
of SCET where those massive gauge bosons, together with the top quark and the Higgs boson, are integrated out. Following
Ref.~\cite{Chiu:2009mg} we denote the theory below $\mu_l$ as
SCET$_{\gamma}$, and the theory above $\mu_l$ as SCET$_{\rm EW}$. The final step (iv) consists of the running from
$\mu_l$ to the factorization scale $\mu_f$. Steps (i) and (ii) are
independent of the masses of the gauge bosons, and of the pattern of
electroweak symmetry breaking, and can be performed in the unbroken
theory, with massless particles. The jet and soft functions are then defined in SCET$_\gamma$ and only contain photon, gluon
and light-fermion radiation. We denote by $\mu_j$ and $\mu_s$ the
scales where the jet and soft functions are defined,
respectively. This whole setup is illustrated in
Fig.~\ref{fig:stepsRGev}. In this paper, we study in
detail the importance of electroweak corrections to the different
ingredients of the factorization formula and
discuss the best way to set the different factorization scales when
both QCD and electroweak corrections are included.

Before going on, a comment regarding the definition of the observable we are considering
is in order, which is no longer unambiguous once electroweak corrections are included. In particular, we need to clarify what we mean by a jet. In the full SM, when one computes
electroweak corrections to $V$+jet one will encounter real radiation
diagrams where there is a photon and a gluon or quark in the
final state. The $p_T$ of the electroweak boson $V$ can be balanced by both the
recoiling parton and the photon. As a consequence, the singularities
of the real-emission diagrams cancel in part with virtual electroweak
corrections to $V$+jet but also in part with QCD corrections to the
$V+\gamma$ process. One should therefore either put some cut which excludes configurations where the $p_T$ of the $V$ is compensated by a hard photon, or consider a more
inclusive observable and include also the $V+\gamma$ process with its QCD corrections. Either
option amounts to a well-defined  observable, and both of them were discussed
in the literature~\cite{Hollik:2007sq,Kuhn:2007cv}. In addition, Refs.~\cite{Denner:2009gj,Denner:2011vu} present electroweak corrections including the leptonic $W$ and $Z$ decay. Comparisons between the different
results seem to indicate that the the size of the corrections is very similar in the two cases  \cite{Kuhn:2007cv}. Here, we consider single-boson production near threshold, where the factorization formula  Eq.~(\ref{eq:schfact}) for the partonic cross section is valid. Since we consider inclusive $V$ production in the threshold limit, the real radiation is encoded in the soft and jet functions, with no phase-space for additional hard radiation.  At leading order in the power counting in the effective theory one will have operators which, in addition to the vector boson $V$, involve: (i) a collinear quark or
gluon field, or (ii) a collinear photon in the final state. Obviously the first operators
give $V$+jet, while the second ones give $V+\gamma$ at Born
level. The different operators do not mix, and we
can consider them separately. We will not include operator (ii) in the following.  The quark (and gluon) jet functions contain collinear
photons, but soft quark radiation is power suppressed in the threshold limit and a quark jet function will not lead to contributions where a photon carries all the energy.
Therefore by considering the threshold limit, and writing down the factorized formula in the
effective theory, we avoid the need to introduce an explicit cut to exclude a hard photon. The cut would affect power suppressed terms, whose size will govern its importance.

The rest of the paper is organized as follows. In Sec.~\ref{sec:prelim}, we first
we describe the kinematics of the process and specify the power
counting that we use. We then give general expressions for the anomalous dimensions and the matching corrections and discuss how the result for specific channels can be obtained.  Section~\ref{sec:Z} contains the results and
plots for $Z$ and $\gamma$ production, while Sec.~\ref{sec:W} contains
the ones for $W$ bosons. In Sec.~\ref{sec:disc} we 
discuss the size of the electroweak corrections, compare
with results from the literature and conclude. Appendix~\ref{sec:appbSM} collects the
beta functions that enter in our results.

\section{Preliminaries\label{sec:prelim}}

\subsection{Kinematics and power counting\label{sec:kinpc}}
There are two partonic channels that are relevant for single
electroweak boson production at leading order: the Compton channel $q\,g\to q\,V$, and the
annihilation channel $q\,\bar{q}\to g\,V$, plus permutations of the
initial-state partons or interchange of $q$ and $\bar{q}$. At
next-to-next-to-leading order also the channel $g\,g\to g\,V$ contributes, but it is only relevant for N$^3$LL accuracy, and we don't
need to consider
it here. The partonic Mandelstam
variables for $a\,b\to c\,V$ are given by $\hat{s}=(p_a+p_b)^2$, $\hat{t}=(p_a-p_V)^2$,
and $\hat{u}=(p_b-p_V)^2$. Throughout the paper, a hat denotes a
partonic quantity.

We now define a counting to be able to specify which terms in the amplitude will be kept
in our results. Defining $a$ as the counting parameter, we use
\begin{equation}
\alpha_s\sim a\quad;\quad L\defeq\log\frac{p_T^2}{M_Z^2}\sim\frac{1}{a}\quad;\quad
\alpha_i\sim a ^2,
\end{equation}
where $\alpha_s$ is the strong coupling and $\alpha_i$ is the ${SU(2)}_L$ or ${U(1)}_Y$
coupling ($\alpha_2$ or $\alpha_1$, respectively; or the electromagnetic coupling $\alpha_{em}$, if we are
in SCET$_{\gamma}$). As was done in
Ref.~\cite{Chiu:2008vv}, we find it convenient to present a table with
the different terms entering in the amplitude. If we denote the amplitude by
$\mathcal{M}$, we have, schematically, that the log of the amplitude
will contain the following terms
\begin{equation}\label{eq:tabterms}
\log \mathcal{M}\sim
\left( \begin{array}{cccc}
\alpha_sL^2+\alpha_iL^2 & \alpha_sL+\alpha_iL & \alpha_s+\alpha_i & \\
\sim\frac{1}{a}+1 & \sim 1+a & \sim a
+a^2 & \\
\\
\alpha_s^2L^3+\alpha_i^2L^3 & \alpha_s^2L^2+\alpha_i^2L^2 &
\alpha_s^2L+\alpha_s\alpha_iL+\alpha_i^2L & \alpha_s^2+\alpha_s\alpha_i+\alpha_i^2\\
\sim\frac{1}{a}+a & \sim 1+a^2 & \sim a +a^2+a^3 & \sim a^2+a^3+a^4\\
\\
\alpha_s^3L^4+\alpha_i^3L^4 &
\alpha_s^3L^3+\alpha_s^2\alpha_iL^3+\alpha_s\alpha_i^2L^3+\alpha_i^3L^3 & \vdots &  \ddots\\
\sim\frac{1}{a}+a^2 &
\sim 1+a+a^2+a^3 & &  \\
\\
\vdots & \vdots & & \\
\end{array}\right),
\end{equation}
where it is understood that the amplitude is normalized such that tree level corresponds to $\mathcal{M}=1$. To obtain the above table,
one needs to take into account that the $\beta$-functions in the SM contain terms
involving couplings from different gauge groups starting at two loops,
and that the cusp anomalous dimension contains terms that mix the
couplings starting at four loops. For the pure-QCD terms, N$^{k-1}$LL accuracy corresponds to keeping all
the terms in the first $k$ columns of the table of
Eq.(\ref{eq:tabterms}). Pure-QCD terms were considered at N$^2$LL
accuracy in Refs.~\cite{Becher:2011fc,Becher:2012xr,Becher:2009th},
here we will also consider the rest of the terms, which involve at least
one $\alpha_i$.

\subsection{Renormalization-group evolution of the hard function\label{sec:anomalous}}

In the effective theory, the resummation is performed by solving
renormalization-group (RG) equations for the hard, jet and soft functions and evolving them to a common scale. To set the stage for the explicit expressions given in the next sections, we now give general expressions for the necessary hard anomalous dimensions and the solution of the associated RG equations. The hard functions are given by renormalized on-shell amplitudes. For massless particles and at the one-loop level, their evolution is governed by the anomalous dimension \cite{Becher:2009cu,Becher:2009qa}
\begin{equation}\label{magic}
   \bm{\Gamma}(\{\underline{p}\},\mu) 
   = \sum_{i < j}\,\frac{\alpha}{4\pi}\,\bm{T}_i\cdot\bm{T}_j\,\Gamma_0\,\ln\frac{\mu^2}{-s_{ij}} 
    + \sum_i\,\,\frac{\alpha}{4\pi} \gamma_0^i \,,
\end{equation}
where $\{\underline{p}\}$ represents the set of momentum vectors of
the external particles, $s_{ij}\defeq2\sigma_{ij}\,p_i\cdot p_j+i0$, and
the sign factor $\sigma_{ij}=+1$ if the momenta $p_i$ and $p_j$ are
both incoming or outgoing, and $\sigma_{ij}=-1$ otherwise. The product
$\bm{T}_i\cdot\bm{T}_j = \sum_a \bm{T}_i^a \bm{T}_j^a $, where
$\bm{T}_i^a$ are the gauge-theory generators in the representation
relevant for particle $i$ (see e.g.\ Ref.~\cite{Becher:2009qa} for more
details). This expression is valid for a general unbroken gauge theory
with coupling constant $\alpha$. The one-loop cusp anomalous dimension
is $\Gamma_0=4$ and the collinear anomalous dimension $ \gamma_0^i$
depends on the representation and the spin of the particle. For a fermion, one has
$\gamma_0^q=-3C_F$, while it takes the value $\gamma_0^g=-\beta_0$ for
a
gauge boson, where it is understood that one uses the Casimir $C_F$
and first beta function coefficient $\beta_0$ that are appropriate
for the corresponding gauge group. Because we count $\alpha_1 \sim
\alpha_2 \sim a^2$, the one-loop expression for the anomalous
dimension is sufficient for the electroweak corrections at order $a$. However, we note that there
are strong all-order constraints on the anomalous dimension
\cite{Becher:2009cu,Gardi:2009qi,Becher:2009qa,Becher:2009qa,Dixon:2009ur,DelDuca:2011ae,Ahrens:2012qz},
which imply in our case that the structure of the anomalous dimension remains the same up to, at least, three-loop order. 

The relevant theory for the hard functions at $\mu \sim p_T \gg M_V$ is the SM in the unbroken phase, and one has to replace
\begin{equation}
\frac{\alpha}{4\pi}\,\bm{T}_i\cdot\bm{T}_j \to
\frac{\alpha_1}{4\pi}\,Y_i\cdot Y_j + \frac{\alpha_2}{4\pi}\, \bm{t}_i
\cdot \bm{t}_j + \frac{\alpha_s}{4\pi}\, \bm{T}_i \cdot \bm{T}_j\, ,
\end{equation}
in the above expression, Eq.~(\ref{magic}), for the anomalous dimension. Here $Y_i$ is the hypercharge of particle $i$, and $\bm{t}_i$ and $\bm{T}_i$ the generators associated with ${SU(2)}_L$ and ${SU(3)}_C$, in the appropriate representation for particle $i$. The relevant theory for the running below $\mu_l$ is QCD+QED and the anomalous dimensions are obtained by substituting
\begin{equation}
\frac{\alpha}{4\pi}\,\bm{T}_i\cdot\bm{T}_j \to
\frac{\alpha_{em}}{4\pi}\,Q_i\cdot Q_j + \frac{\alpha_s}{4\pi}\,
\bm{T}_i \cdot \bm{T}_j\, ,
\end{equation}
in the general expression, where $Q_i$ is the electric charge of
particle $i$.

In our case, the hard amplitudes are $q\,\bar{q} \to g \, V$ and crossings thereof. They involve only three charged particles under any of the gauge groups. In the three-particle case, charge conservation $\sum_i \bm{T}_i = 0$ can be used to rewrite the color-dipole terms in the form
\begin{align}
2\sum_{i < j}\,\bm{T}_i\cdot\bm{T}_j\,\ln\frac{\mu^2}{-s_{ij}} 
   &= - (C_1+C_2-C_3)\,\ln\frac{\mu^2}{-s_{12}} - (C_1+C_3-C_2)\,\ln\frac{\mu^2}{-s_{13}} -(C_2+C_3-C_1)\,\ln\frac{\mu^2}{-s_{23}} \,,
   \end{align}
where $C_i = \bm{T}_i\cdot\bm{T}_i$ is the quadratic Casimir operator associated with leg $i$. This makes it clear that the anomalous dimensions for vector-boson production are diagonal in gauge-group space.

Below, we will write the RG equation for the hard function for production of a vector boson $V$ in the partonic channel $a \,b \to c\, V$ in the form
\begin{equation}
\frac{d}{d\ln\mu} H_{ab,V}(\hat{s},\hat{t},\mu) = \left[ \Gamma^{(V)} \ln\frac{\hat{s}}{\mu^2} +\gamma^{(V)}_{ab} \right]  \,H_{ab,V}(\hat{s},\hat{t},\mu)\,,
\end{equation}
where we suppress the dependence of the non-cusp anomalous dimension $\gamma^{(V)}_{ab}$ on $\hat{s}$ and $\hat{t}$. We write the solution to this equation in terms of an evolution factor times the hard function at a high scale $\nu$:
\begin{equation}
H_{ab,V}(\hat{s},\hat{t},\mu) = \mathcal{U}_{ab}^{(V)}\left(\nu,\mu\right)\, H_{ab,V}(\hat{s},\hat{t},\nu)\,.
\end{equation}
At the order we are working, we do not need to consider electroweak
corrections to the matching, and the functions
$H_{ab,V}(\hat{s},\hat{t},\mu_h)$ are then given by the corresponding
QCD results. We thus do not
need to distinguish the ${U(1)}_Y$ hard function
$H_{ab,B}(\hat{s},\hat{t},\mu_h)$ from the $SU(2)_L$ hard function $H_{ab,W^3}(\hat{s},\hat{t},\mu_h)$ and will denote the common QCD hard function for $Z$-boson production simply by $H_{ab,Z}(\hat{s},\hat{t},M_Z,\mu_h)$. We furthermore keep the power-suppressed $M_Z$ dependence both in
  $H_{q\bar{q},Z}(\hat{s},\hat{t},M_Z,\mu_h)$ and in the Born-level cross sections.

The pure-QCD hard functions and their evolution were given in \cite{Becher:2009th,Becher:2012xr}. The contribution of the electroweak gauge coupling $\alpha_i$ to the evolution factor has the following form:
\begin{equation}\label{eq:defRD}
\left. \ln \mathcal{U}_{ab}^{(V)}\left(\nu,\mu\right)\right|_{\alpha_i}= 2 S ^{(V)}(\nu,\mu)-A _{ab}^{(V)}(\nu,\mu)- \ln\frac{\hat{s}}{\nu^2}\,A_{\Gamma}^{(V)}(\nu,\mu) +2S _{\alpha_s\alpha_i}(\nu,\mu)\,,
\end{equation}
where the functions $S ^{(V)}(\nu,\mu)$ and $A _{ab}^{(V)}(\nu,\mu)$ are \cite{Becher:2009th,Becher:2006mr}
\begin{equation}\label{eq:defSA}
S^{(V)}(\nu,\mu)=-\int_{\alpha_i(\nu)}^{\alpha_i(\mu)}d\alpha\frac{\Gamma^{(V)}(\alpha)}{\beta(\alpha)}\int_{\alpha_i(\nu)}^\alpha\frac{d\alpha'}{\beta(\alpha')}\quad;\quad A_{ab}^{(V)}(\nu,\mu)=-\int_{\alpha_i(\nu)}^{\alpha_i(\mu)}d\alpha\frac{\gamma_{ab}^{(V)}(\alpha)}{\beta(\alpha)}\,.
\end{equation}
The function $A _{\Gamma}^{(V)}(\nu,\mu)$ is obtained by replacing $\gamma^{(V)}_{ab}$ with $\Gamma^{(V)}$ in $A _{ab}^{(V)}(\nu,\mu)$. It is understood that the
appropriate anomalous dimension and beta function for each coupling is used in
Eq.~(\ref{eq:defSA}). The explicit expressions for the beta function
coefficients appearing throughout the paper are
collected in Appendix~\ref{sec:appbSM}. In addition to the contributions from the individual gauge groups, we also need to take into account mixing terms at the order we are working. These are encoded in the last term in Eq.~(\ref{eq:defRD}). The mixing contribution is $S
_{\alpha_s\alpha_{1,2}}$ when $V=W^{1,2,3}$ or $B$, and $S
_{\alpha_s\alpha_{em}}$ when $V=Z$ ,$W^{\pm}$, or $\gamma$, for the hard  running below the scale $\mu_l$. It is given by the function
$S(\nu,\mu)$ for the strong coupling keeping only the terms of order $a$
that contain one $\alpha_i$, which come from the expansion of the beta
function for $\alpha_s$ in the denominators of Eq.~(\ref{eq:defSA}).
These corrections correspond to the terms with one $\alpha_i$ and an arbitrary number
of $\alpha_s$ in the second column of Eq.~(\ref{eq:tabterms}), all of
which are of order $a$. For the second stage of hard running, below the scale $\mu_l$, the logarithm $\ln\hat{s}/\mu_l^2$ arising in Eq.~(\ref{eq:defRD}) is large because $\mu_l^2\ll \hat{s}$. In this case, we also need to include a mixing term $A_{\alpha_s\alpha_{em}}$ arising in the function $A_{\Gamma}^{(V)}$.

\subsection{Collinear factorization anomaly\label{sec:collinear}}

For massive Sudakov problems, the individual collinear and soft diagrams are not well defined and need additional regularization beyond the standard dimensional regularization. This can be done with an analytic regulator \cite{Smirnov:1997gx}. The additional regulator can be removed once the contributions from the individual collinear regions are combined, but as a result a large logarithm arises in the matching of SCET$_{\rm EW}$ to SCET$_{\rm \gamma}$ \cite{Chiu:2007yn}. The presence of large logarithms in the matching is problematic since such logarithms are not generated by RG evolution but need to be resummed. This collinear anomaly also arises in many other observables, in particular in transverse-momentum dependent quantities \cite{Becher:2010tm}. It was shown in \cite{Chiu:2007dg} that the additional logarithm exponentiate. The exponentiation is derived from the requirement that the regulator dependence must cancel among the different collinear and soft pieces \cite{Chiu:2007dg,Becher:2010tm}, or alternatively, from solving an evolution equation in the associated regulator scale \cite{Chiu:2012ir}.

While the standard electroweak matching corrections are beyond our
accuracy, the logarithmically enhanced pieces due to the collinear anomaly need to be included. One-loop collinear functions for the Standard Model were given in \cite{Chiu:2009ft}. The logarithmically enhanced piece has the general form
\begin{equation}
D^i_C = \frac{\alpha}{4\pi} \, \frac{\Gamma_0}{4}\,\bm{T}_i\cdot \bm{T}_i  \ln \frac{M^2}{\mu^2} \ln\frac{\hat{s}}{\mu^2}\,,
\end{equation}
where $\alpha$ stands for the coupling of the broken gauge group whose boson has a mass $M$. The full one-loop expression involves a sum over pairs like Eq.~(\ref{magic}), but since we only need the leading logarithmic contribution, we have replaced $s_{ij}\to -\hat{s}$ and have used 
charge conservation $\sum_i \bm{T}_i = 0$ to write it in the above form. These collinear functions need to be computed in the broken phase and to obtain them, one replaces \cite{Chiu:2009ft}
\begin{equation}
\alpha\, \bm{T}_i\cdot \bm{T}_i \to \alpha_W\,(\bm{t}_i\cdot \bm{t}_i - (\bm{t}_i^3)^2) + \alpha_Z\, (\bm{t}_i^Z)^2\,,
\end{equation}
where $\bm{t}^Z = \bm{t}^3 - s_W^2 Q$, $\alpha_W=\alpha_2$, and
$\alpha_Z = \alpha_1/s_W^2 =  \alpha_2/c_W^2 =
\alpha_{em}/(c_W^2\,s_W^2)$, with
$c_W\defeq\cos \theta_W$, $s_W\defeq\sin \theta_W$, and $\theta_W$ the
weak-mixing angle. To obtain the eigenvalues of the Casimir operators acting on the $W$-boson fields, we need to work with the generators in the adjoint representation. We find
\begin{align}
(\bm{t}\cdot \bm{t} - (\bm{t}^3)^2) | W^\pm\rangle &= | W^\pm\rangle & (\bm{t}\cdot \bm{t} - (\bm{t}^3)^2) | W^3 \rangle &= 2 | W^3\rangle \nonumber &
 (\bm{t}^Z)^2 | W^\pm\rangle &= c_W^4 | W^\pm\rangle &(\bm{t}^Z)^2 | W^3 \rangle &= 0\,.
\end{align}
We will use the notation
\begin{equation}
D_{q\bar{q}}^{(W^3\to Z)}(\mu) =  D^q_C  + D^{\bar{q}}_C + D^{W^3}_C 
\end{equation}
for the collinear function for $Z$-boson production arising from the
operator with field content $q\bar{q} W^3$ (and analogous notations
for the rest of the collinear functions). As we stated above, this contribution exponentiates, so the relevant factor in the cross section is
\begin{equation}
\mathcal{D}_{q\bar{q}}^{(W^3\to Z)}(\mu) = e^{D_{q\bar{q}}^{(W\to Z)}(\mu)}\,.
\end{equation}

\section{\boldmath Results for $Z$ and $\gamma$ production}\label{sec:Z}

We now give the results for $Z$ production and will afterwards discuss how they must be modified to also obtain the cross section for $\gamma$ production.
The hadronic cross section for the $Z$-boson case is given by
\begin{equation}\label{eq:hadrcs}
\frac{d\sigma_Z}{dp_Tdy}=2p_T\sum_{ab}\int
dx_1dx_2f_a(x_1)f_b(x_2)\left[\hat{s}\frac{d\hat{\sigma}_{ab,Z}}{d\hat{s}d\hat{t}}\right],
\end{equation}
where $f_a(x)$ is the PDF for parton $a$. The sum runs over
$a,b=q,\bar{q},g$, and $y$ is the $Z$'s rapidity. The factorized form
for the partonic cross section for the channel $a\,b\to c\,Z$ is 
\cite{Becher:2009th,Becher:2011fc}
\begin{equation}\label{eq:factpartZ}
\hat{s}\frac{d\hat{\sigma}_{ab,Z}}{d\hat{s}d\hat{t}}=\hat{\sigma}^B_{ab,Z}H_{ab,Z}(\hat{s},\hat{t},M_Z,\mu) \int dk\,J_c(M_X^2-2E_Jk,\mu)\,S_{ab,Z}(k,\mu),
\end{equation}
where $M_X^2=(p_a+p_b-p_Z)^2$, and $E_J$ is the energy of the jet. At
Born level $J_c(p^2,\mu)$ and $S_{ab,Z}(k,\mu)$ reduce to delta functions of
their first arguments, and the hard function $H_{ab,Z}$ is equal to 1. The
Born-level cross sections are given by
\begin{equation}
\hat{\sigma}^B_{q\bar{q},Z}=\frac{\pi\alpha_{em}\alpha_s}{\hat{s}}\frac{2C_F}{N_c}(I^Z)^2\left(\frac{\hat{t}^2+\hat{u}^2+2M_Z^2\hat{s}}{\hat{t}\hat{u}}\right) \quad ;\quad \hat{\sigma}^B_{qg,Z}=-\frac{\pi\alpha_{em}\alpha_s}{\hat{s}}\frac{1}{N_c}(I^Z)^2\left(\frac{\hat{t}^2+\hat{s}^2+2M_Z^2\hat{u}}{\hat{t}\hat{s}}\right) ,
\end{equation}
with $C_F=(N_c^2-1)/(2N_c)$, $N_c$ the number of colors, and
$I^Z\defeq\left(\frac{c_W}{s_W}t^3-\frac{s_W}{c_W}Y\right)$, with $t^3$ the weak isospin and $Y$ the hypercharge (the electric charge $Q$ is
given by $Q=t^3+Y$). Note that we keep the $M_Z$ terms in the
Born-level cross section, and in kinematical factors.

The $Z$ boson in the low-energy broken theory can come from the
$U(1)_Y$ gauge boson $B$ or the $SU(2)_L$ gauge boson $W^3$ in the
unbroken theory. We therefore need to consider the amplitudes
$a\,b\to c\,W^3$ and $a\,b\to c\,B$ in the high-energy unbroken theory, and combine them according to
$Z=c_WW^3-s_WB$, which can be thought of as part of
the tree-level matching condition at $\mu_l$. For the pure-QCD terms, the log
resummation is the same for the $B$ and $W^3$ terms, and therefore the
amplitude is still proportional to $I^Z$ after resummation, but when
we include electroweak corrections the $B$ and $W^3$ terms receive
different contributions and the resummed amplitude is no longer
proportional to $I^Z$. The external $Z$, and the external $W$ in the next section, are treated with a boosted version of a Heavy Quark Effective Theory (HQET) field in SCET$_{\gamma}$ 
\cite{Chiu:2009mg,Fleming:2007qr} (we use the standard name HQET, despite the fact that the heavy particle is not a quark in our case). Since the $Z$ is neutral, its jet
function $J_Z$ is trivial (i.e. exactly a $\delta$ function), and it
does not appear explicitly in the factorization formula in
Eq.~(\ref{eq:factpartZ}). That is, we have already integrated over the
associated convolution variable that would appear in Eq.~(\ref{eq:schfact}).

\subsection{Cross section in SCET}
In this section we present the results for the different ingredients
that enter in the factorized cross section formula in SCET, Eqs.~(\ref{eq:hadrcs})-(\ref{eq:factpartZ}). The hard function (times Born-level cross section) for the
annihilation channel is given by
\[
\hat{\sigma}^B_{q\bar{q},Z}H_{q\bar{q},Z}(\hat{s},\hat{t},M_Z,\mu_f)=\frac{\pi\alpha_s(\mu_h)}{\hat{s}}\frac{2C_F}{N_c}\left(\frac{\hat{t}^2+\hat{u}^2+2M_Z^2\hat{s}}{\hat{t}\hat{u}}\right)\left|\mathcal{U}_{q\bar{q}}^{(Z)}\left(\mu_l,\mu_f\right)\right|^2
\]
\begin{equation}\label{eq:evHqq}
\times\left|c_W\sqrt{\alpha_2(\mu_h)}t^3 \mathcal{D}_{q\bar{q}}^{(W^3\to Z)}\left(\mu_l\right)\mathcal{U}_{q\bar{q}}^{(W^3)}\left(\mu_h,\mu_l\right)-s_W\sqrt{\alpha_1(\mu_h)}Y\mathcal{D}_{q\bar{q}}^{(B\to Z)}\left(\mu_l\right)\mathcal{U}_{q\bar{q}}^{(B)}\left(\mu_h,\mu_l\right)\right|^2 H_{q\bar{q},Z}(\hat{s},\hat{t},M_Z,\mu_h).
\end{equation}

In Eq.~(\ref{eq:evHqq}), the $\mathcal{U}_{q\bar{q}}^{(V)}(\mu_h,\mu_l)$ factors
encode the running from $\mu_h$ to $\mu_l$ in SCET$_{EW}$, and
$\mathcal{U}_{q\bar{q}}^{(Z)}(\mu_l,\mu_f)$ the running from $\mu_l$ to
$\mu_f$ in SCET$_{\gamma}$. The general structure of the anomalous
dimensions was given in Section \ref{sec:anomalous}, we present in the following the expressions needed to account for the terms up to
order $a$ in Eq.~(\ref{eq:tabterms}). The explicit results for the electroweak part of the anomalous dimensions we need are
\begin{eqnarray}\label{eq:gammWS}
\Gamma^{(W^3)} & = & \frac{\alpha_1}{4\pi}Y^2
\Gamma_0+\frac{\alpha_2}{4\pi}\frac{7}{4}\Gamma_0, \nonumber \\
\gamma_{q\bar{q}}^{(W^3)} & = & \frac{\alpha_1}{4\pi}Y^2\left( -6- i\pi
  \Gamma_0\right)+\frac{\alpha_2}{4\pi}\frac{1}{4}\left(-18-4\beta_0^{\alpha_2}+ i\pi \Gamma_0-4\Gamma_0 \log\frac{\hat{s}^2}{\hat{t}\hat{u}}\right),  \nonumber\\
\Gamma^{(B)} & = & \frac{\alpha_1}{4\pi}Y^2
\Gamma_0+\delta_{\lambda L}\frac{\alpha_2}{4\pi}\frac{3}{4}\Gamma_0,\\
\gamma_{q\bar{q}}^{(B)} & = & \frac{\alpha_1}{4\pi}\left[Y^2\left( -6-i\pi
    \Gamma_0\right)-\beta_0^{\alpha_1}\right]+\delta_{\lambda
  L}\frac{\alpha_2}{4\pi}\frac{3}{4}\left(-6-i\pi\Gamma_0\right),  \nonumber\\
\Gamma^{(Z)} & = & \frac{\alpha_{em}}{4\pi}Q^2 \Gamma_0, \nonumber\\
\gamma_{q\bar{q}}^{(Z)} & = & \frac{\alpha_{em}}{4\pi}Q^2\left( -6-i\pi \Gamma_0\right),  \nonumber
\end{eqnarray}
where $\lambda=L,R$ indicates if the quark is left- or right-handed.

The factors $\mathcal{D}_{q\bar{q}}^{(V\to
  B)}(\mu_l)$ encode the matching from SCET$_{EW}$ to SCET$_{\gamma}$
at $\mu_l$. Using the results of Section \ref{sec:collinear}, the relevant functions are obtained as
\begin{equation}
\begin{aligned}
D_{q\bar{q}}^{(W^3\to Z)}(\mu) & = 
\frac{1}{4\pi}2\log\frac{\hat{s}}{\mu^2}\left[\alpha_{em}\left(I^Z\right)^2\log\frac{M_Z^2}{\mu^2}+\alpha_2\frac{3}{2}\log\frac{M_W^2}{\mu^2}\right],\\
D_{q\bar{q}}^{(B\to Z)}(\mu) & = 
\frac{1}{4\pi}2\log\frac{\hat{s}}{\mu^2}\left[\alpha_{em}\left(I^Z\right)^2\log\frac{M_Z^2}{\mu^2}+\delta_{\lambda
    L}\alpha_2\frac{1}{2}\log\frac{M_W^2}{\mu^2}\right].
    \end{aligned}
\end{equation}

The factor $H_{q\bar{q},Z}$ encodes
the matching from the SM to SCET$_{EW}$. At the order we are working
it is 1 plus pure-QCD terms. Here, we do not explicitly show the pure-QCD part of the running and the matching expressions, which were
considered in previous papers, see
Refs.~\cite{Becher:2011fc,Becher:2012xr}. The hard function for the Compton channel is related to
Eq.~(\ref{eq:evHqq}) by crossing.

The leading electroweak corrections for the soft and jet functions are
\begin{equation}\label{eq:softjetLa}
S_{q\bar{q},Z}(k,\mu_f)=e^{-4Q^2S(\mu_s,\mu_f)}\delta(k)\quad;\quad J_q(p^2,\mu_f)=e^{-4Q^2S(\mu_j,\mu_f)}\delta(p^2),
\end{equation}
and $S_{qg,Z}$ and $J_g$ remain delta functions at this
order. $S(\nu,\mu)$ in Eq.~(\ref{eq:softjetLa}) is given by
Eq.~(\ref{eq:defSA}) with
$\Gamma=\frac{\alpha_{em}}{4\pi}\Gamma_0$. As we will discuss in the
next section, we will not include subleading electroweak corrections in
the soft and jet functions for our numerical evaluations, therefore,
we do not write them explicitly.

The results for direct photon production can readily be obtained from the ones for $Z$ production given above. The hard function in the photon case can be obtained from the
corresponding equation for the $Z$ case, Eq.~(\ref{eq:evHqq}), with
the following changes: (i) $M_Z$ should be set to 0 inside the
parenthesis in the first line and in
$H_{q\bar{q},Z}(\hat{s},\hat{t},M_Z,\mu_h)$ (ii) the matching
condition at $\mu_l$ should be changed according to the following substitution:
\begin{equation}
\begin{aligned}
c_W\sqrt{\alpha_2(\mu_h)}t^3 \mathcal{D}_{q\bar{q}}^{(W^3\to
  Z)} & \to & s_W\sqrt{\alpha_2(\mu_h)}t^3 \mathcal{D}_{q\bar{q}}^{(W^3\to
  \gamma)},\\
-s_W\sqrt{\alpha_1(\mu_h)}Y\mathcal{D}_{q\bar{q}}^{(B\to
  Z)} & \to & c_W\sqrt{\alpha_1(\mu_h)}Y\mathcal{D}_{q\bar{q}}^{(B\to
  \gamma)},
  \end{aligned}
\end{equation}
with
$\mathcal{D}_{q\bar{q}}^{(W^3\to\gamma)}=\mathcal{D}_{q\bar{q}}^{(W^3\to
  Z)}$ and
$\mathcal{D}_{q\bar{q}}^{(B\to\gamma)}=\mathcal{D}_{q\bar{q}}^{(B\to
  Z)}$ at the order we are working, and (iii)
$\mathcal{U}_{q\bar{q}}^{(Z)}\left(\mu_l,\mu_f\right)\to
\mathcal{U}_{q\bar{q}}^{(\gamma)}\left(\mu_l,\mu_f\right)$. At the
order we need them, the
anomalous dimensions for $\mathcal{U}_{q\bar{q}}^{(\gamma)}$ are given by
\begin{eqnarray}
\Gamma^{(\gamma)} & = & \Gamma^{(Z)},\\
\gamma_{q\bar{q}}^{(\gamma)} & = & \gamma_{q\bar{q}}^{(Z)}-\frac{\alpha_{em}}{4\pi}\beta_0^{\alpha_{em}}.
\end{eqnarray}

The electroweak corrections to the soft functions are the same as in
the $Z$ case, see Eq.~(\ref{eq:softjetLa}). In principle there
is now a jet function $J_{\gamma}$ for the photon, which contains
contributions from light fermions, analogous to the ones in the QCD
jet function for the gluon. Those were not present in the $Z$
case because there we had an HQET field. As we will discuss below in Sec.~\ref{subsec:scalset}, we do not need to include
these terms. The photon jet function is therefore just a delta
function and we recover the same structure for the factorization
formula that we had in the $Z$ case.

\subsection{Scale setting and numerical results}\label{subsec:scalset}

To evaluate the cross section numerically, we need to set the values of the different scales
that appear in the SCET factorization formula, when both electroweak
and Sudakov corrections are included. We recall
that the electroweak corrections
can be quite significant, around 20\% for $p_T$ around 1 TeV, but the pure-QCD corrections are, of course, also important. The hard, jet, and soft scales that are appropriate for the pure-QCD terms were
determined in Refs.~\cite{Becher:2011fc,Becher:2012xr}, following the
procedure advocated in Ref.~\cite{Becher:2007ty}. They were obtained as
\begin{equation}\label{eq:scalespWZ}
\mu_h=\frac{13p_T+2M_V}{12}-\frac{p_T^2}{\sqrt{s}}\quad;\quad \mu_j=\frac{7p_T+2M_V}{12}\left(1-2\frac{p_T}{\sqrt{s}}\right),
\end{equation}
and $\mu_s=\mu_j^2/\mu_h$. The hard scale in
Eq.~(\ref{eq:scalespWZ}) is of order $p_T$ and therefore also
adequate for the electroweak corrections. On the other hand, to resum
the electroweak Sudakov corrections, we performed a running from $\mu_h$
to the low scale $\mu_l$ in the unbroken gauge theory, and then matched
to a broken gauge theory. We use $\mu_l\sim M_V$, since this is the scale at
which we integrate out the massive gauge bosons. The jet and soft
functions are then defined below the scale $\mu_l$ and contain
only light degrees of freedom, but no $W$ or $Z$ bosons, as is appropriate for
the observable we are studying. The jet and soft scales in
Eq.~(\ref{eq:scalespWZ}), though, are above $M_V$ for values of $p_T$ where the
LHC will measure, and there is thus an apparent difficulty here, in the
sense that the QCD values for the scales are not appropriate for the electroweak
corrections. In practice, this does not lead to problems because
the main part of the electroweak corrections is contained in the hard
function: we have checked that the change in the cross section due to
the leading electroweak corrections of the jet and soft functions is at the level of 1\% or below for the range of $p_T$ we
study. Therefore, we can consider the jet and soft functions just with
leading electroweak corrections; at this order
the strong and electroweak corrections do not mix (see
Eq.~(\ref{eq:softjetLa})), and we can effectively set $\mu_j=M_V$ just
in the electroweak part of the jet function. The alternative to
that would be to choose $\mu_j$ and $\mu_s$ of order $M_V$
everywhere, which would be in accordance with the chain of effective theories we used to resum
the electroweak Sudakov corrections. There is not any obstacle to do
that, but this scale setting would generate larger
uncertainties in the QCD part, and the final result would be less
precise. It is therefore better to ignore the sub-leading, numerically
negligible, electroweak
corrections in the jet and soft functions, and to use the scales in Eq.~(\ref{eq:scalespWZ}) everywhere
except in the electroweak part of the jet and soft functions, where we use
$\mu_j=M_V$, and $\mu_s=\mu_j^2/\mu_h$, accordingly.

Having set the scales, we now present plots of the results for the cross section for $Z$
production. We include electroweak Sudakov as well as QCD corrections.  The default values for the scales $\mu_h$, $\mu_j$, and $\mu_s$ are
fixed according to the discussion above. The default
values for the low-matching scale, $\mu_l$, and the factorization
scale, $\mu_f$, are $\mu_l=\mu_f=M_V$. We will vary these scales by a
factor of 2 to estimate the uncertainties. In all our plots we use the
NNLO MSTW 2008 PDF set \cite{Martin:2009iq}. Note that for
consistency of the factorization formula one should include quantum
electrodynamics (QED) effects in the PDFs. This PDF set, though, does
not include QED effects. There are some older PDF sets that do include
QED corrections \cite{Martin:2004dh}, but these have lower accuracy for the QCD part. Since
the QED corrections in the PDFs should not be very important according
to our discussion above, it is
better to use a newer PDF set with higher QCD orders. The numerical values for
the couplings and masses that we use read $M_Z=91.1876$~GeV,
$M_W=80.399$~GeV, $\alpha_s(M_Z)=0.1171$,
$\alpha_{em}(M_Z)=(127.916)^{-1}$, $\sin^2\theta_W=0.2226$, $V_{ud} = 0.97425$, $V_{us} = 0.22543$, $V_{ub} = 0.00354$, $V_{cd} = 0.22529$,
$V_{cs} = 0.97342$ and $V_{cb} = 0.04128$. We present our results for
the LHC at 7~TeV, for an easier comparison with the results in Ref.~\cite{Becher:2012xr}. The relative size of the corrections is very similar at 13~TeV.

\begin{figure}
\renewcommand{\arraystretch}{2.5}
\begin{tabular}{lr}
\includegraphics[width=8cm]{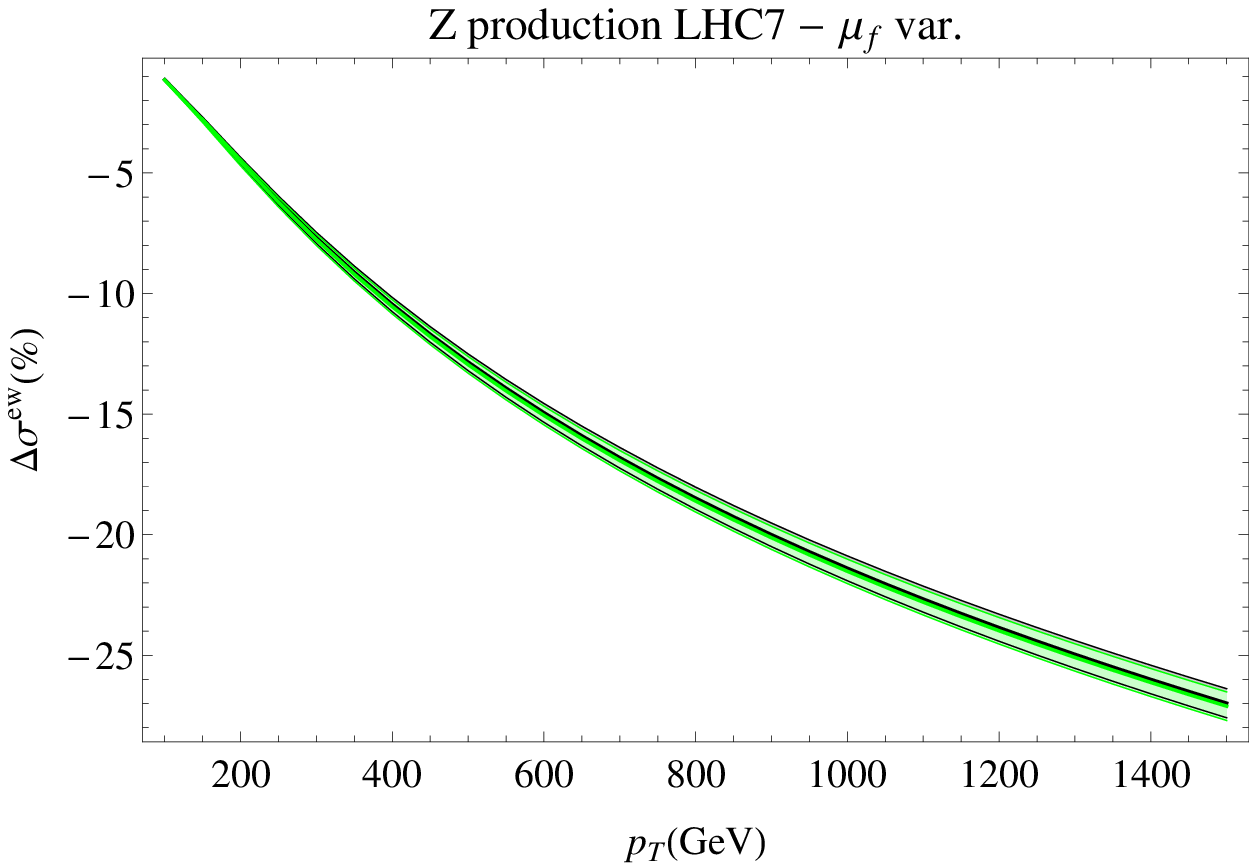}\phantom{bc}&
\includegraphics[width=8cm]{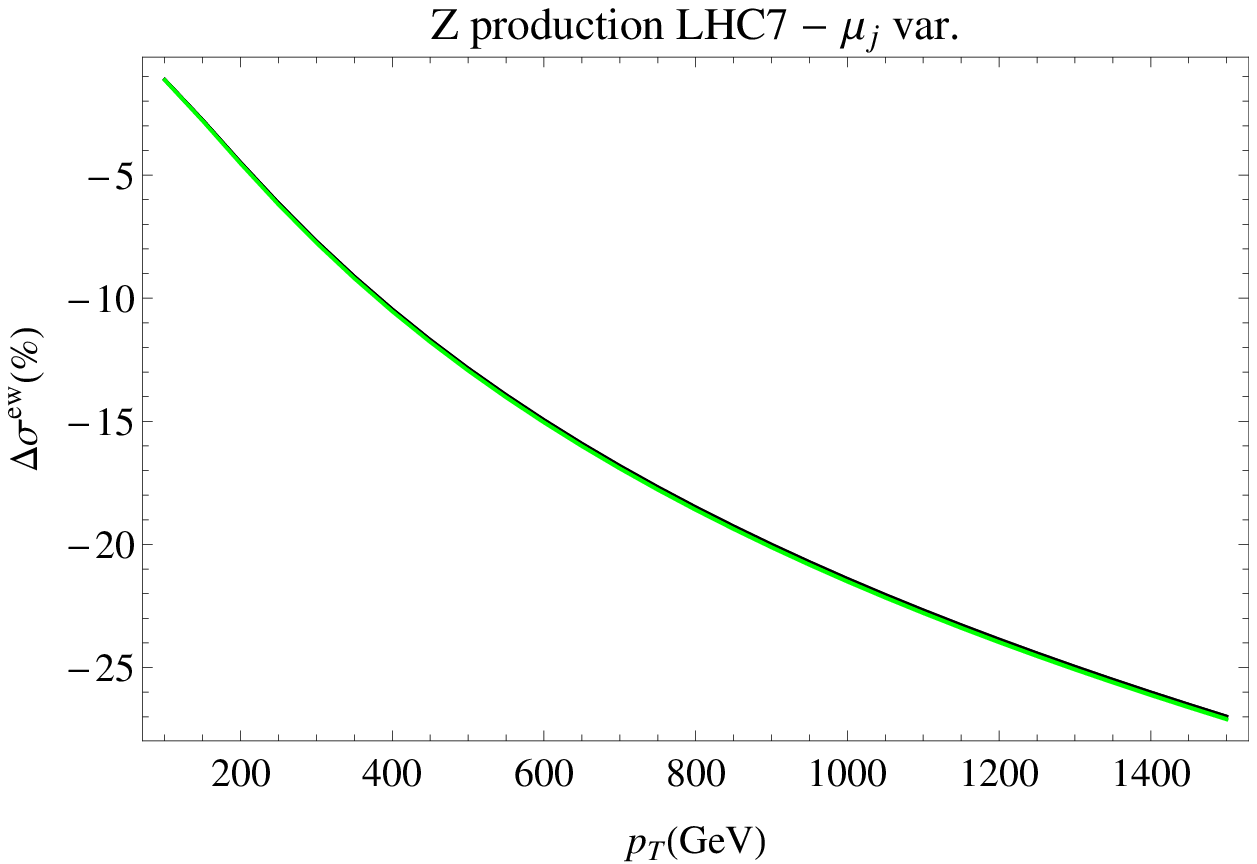} \\
\includegraphics[width=8cm]{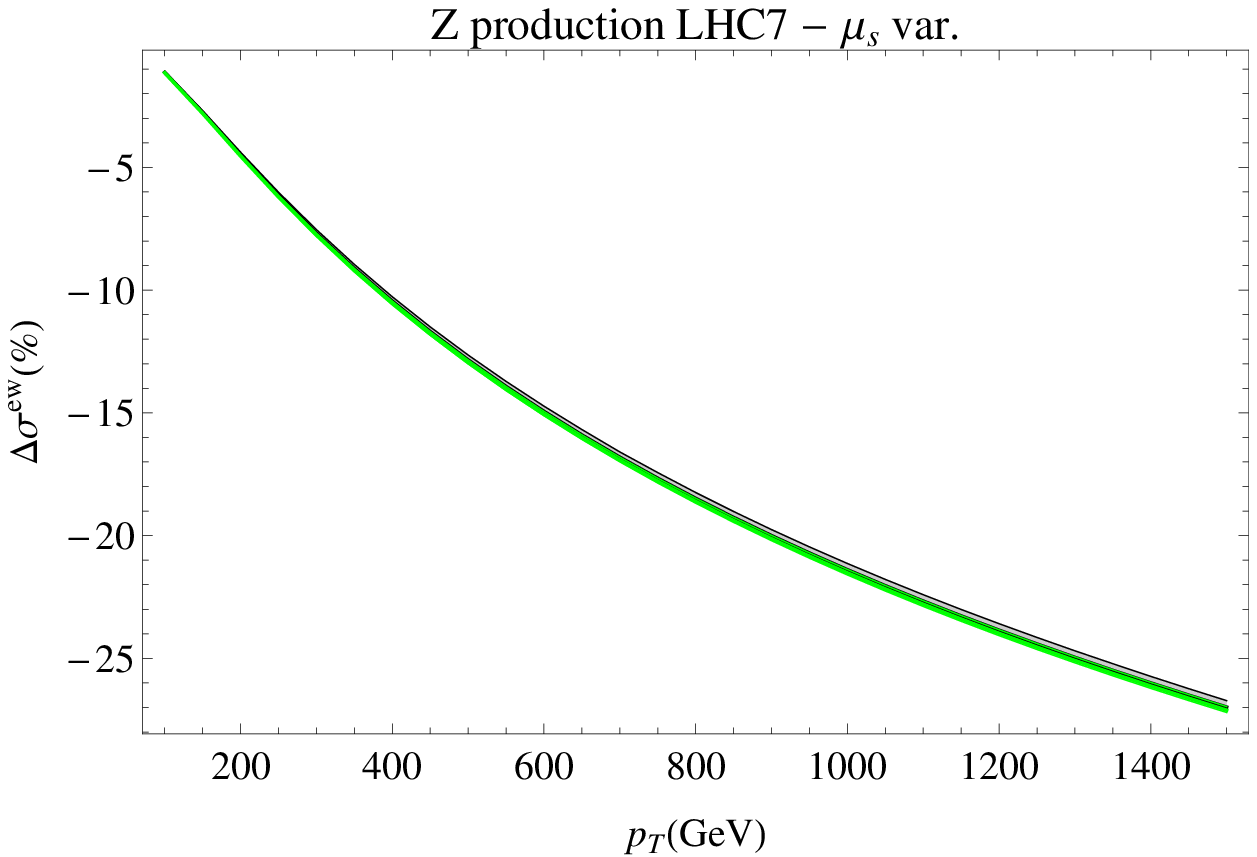}\phantom{bc} &
\includegraphics[width=8cm]{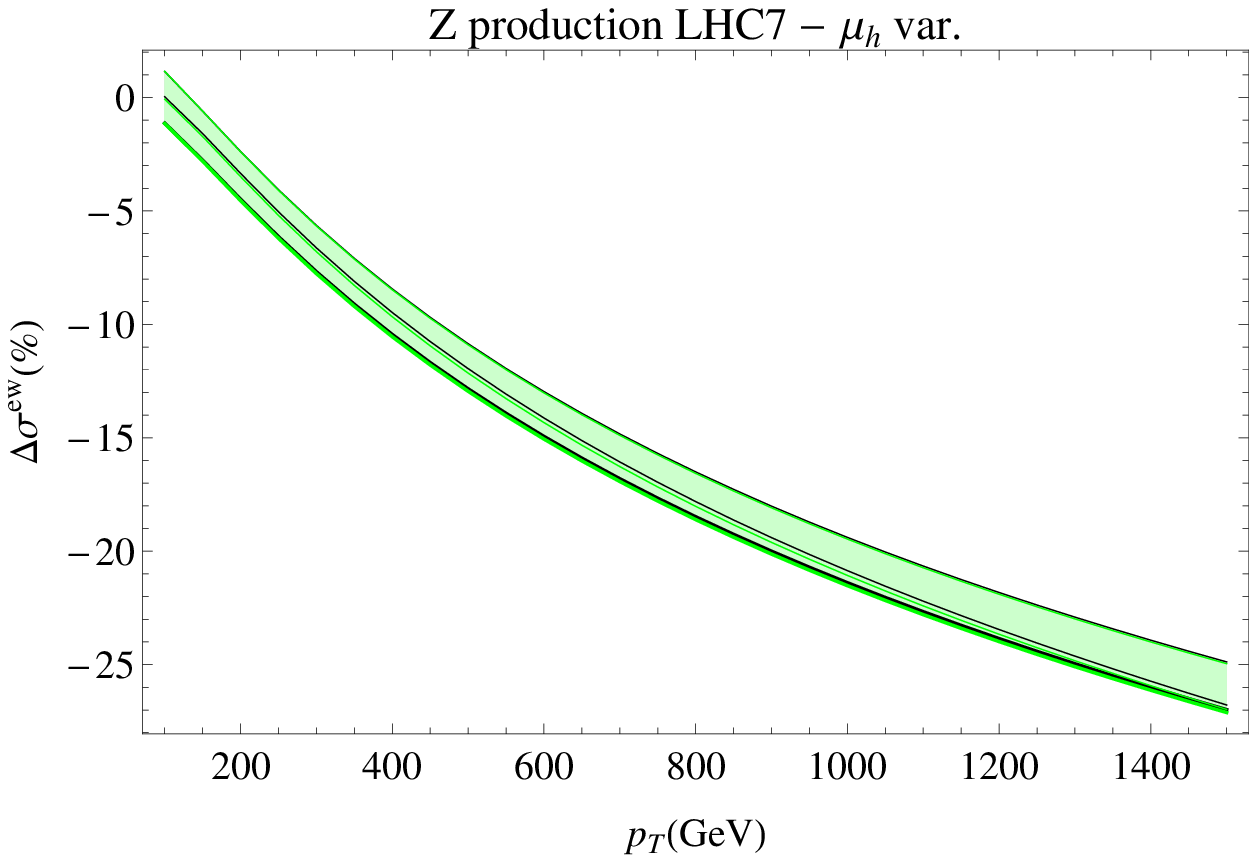} \\
\multicolumn{2}{c}{\includegraphics[width=8cm]{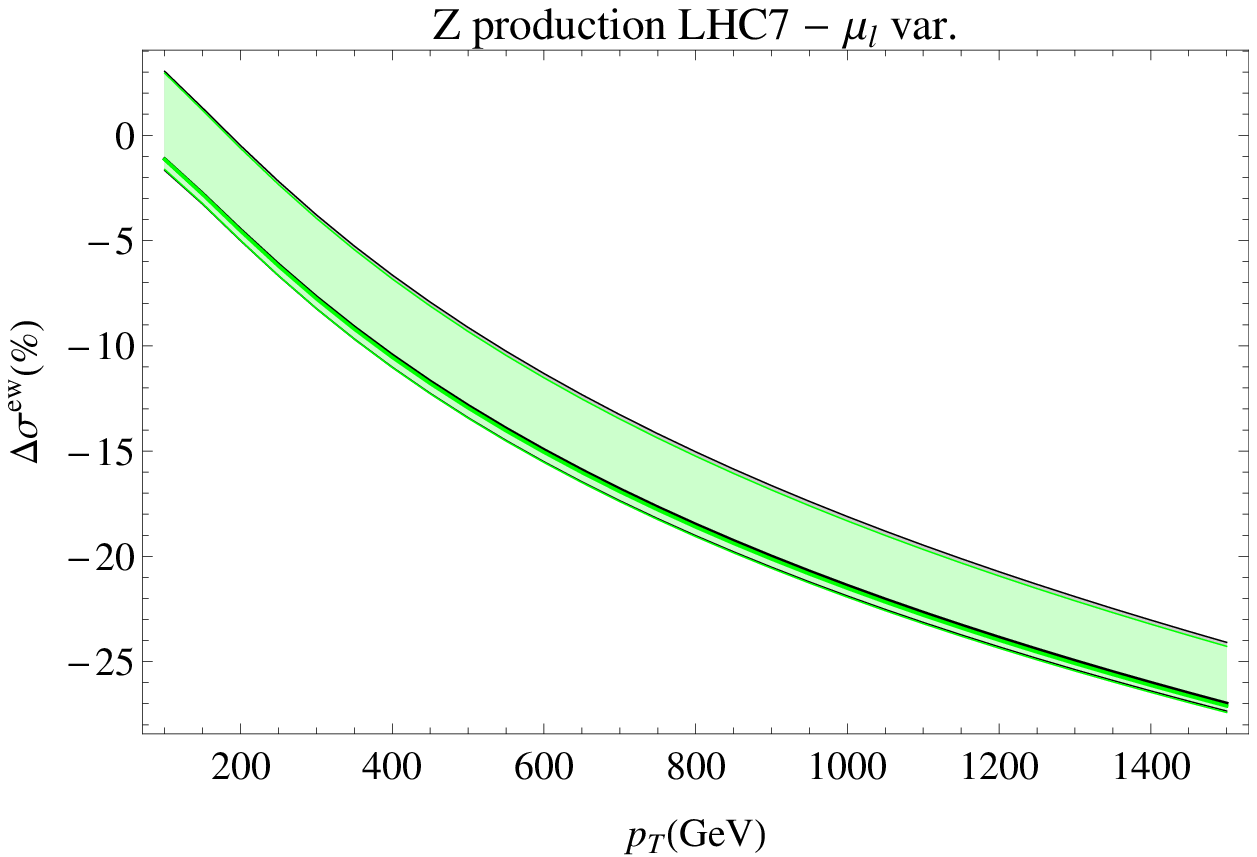}}
\end{tabular}
\caption{$Z$ production for the LHC at 7 TeV. We plot the difference of
  cross sections with and without electroweak corrections, normalized to the QCD result, as defined in Eq.~(\ref{eq:diffsigew}). The black (darker) curves and
  bands correspond to $i=1$, and the green (lighter) ones to $i=2$,
  note that they basically overlap. In each plot we vary the corresponding scale, denoted at the
  top, by a factor of 2.}\label{fig:ZdiffewQCD}
\end{figure}

\begin{figure}
\centering
\includegraphics[width=8cm]{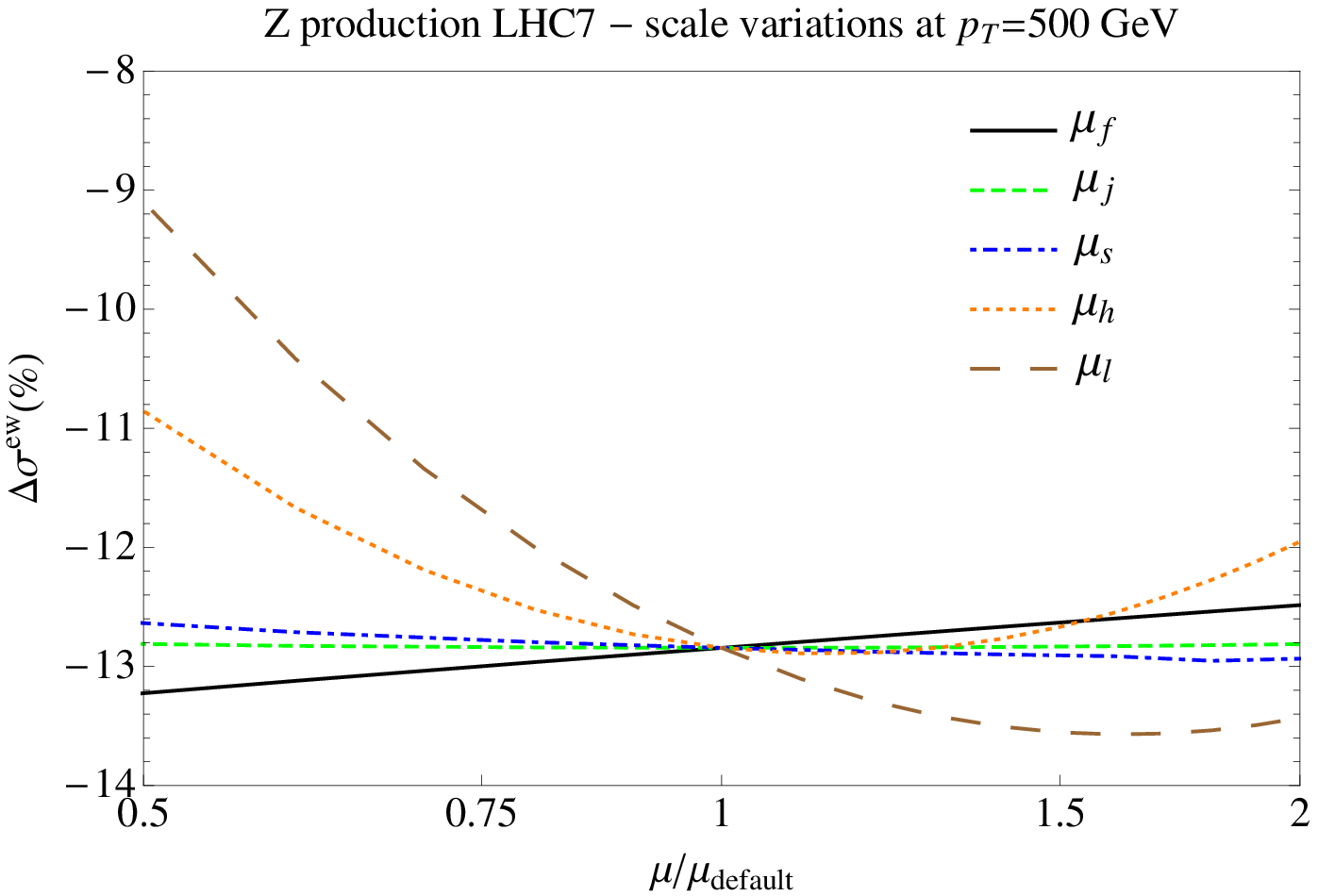}\phantom{ab}
\includegraphics[width=8cm]{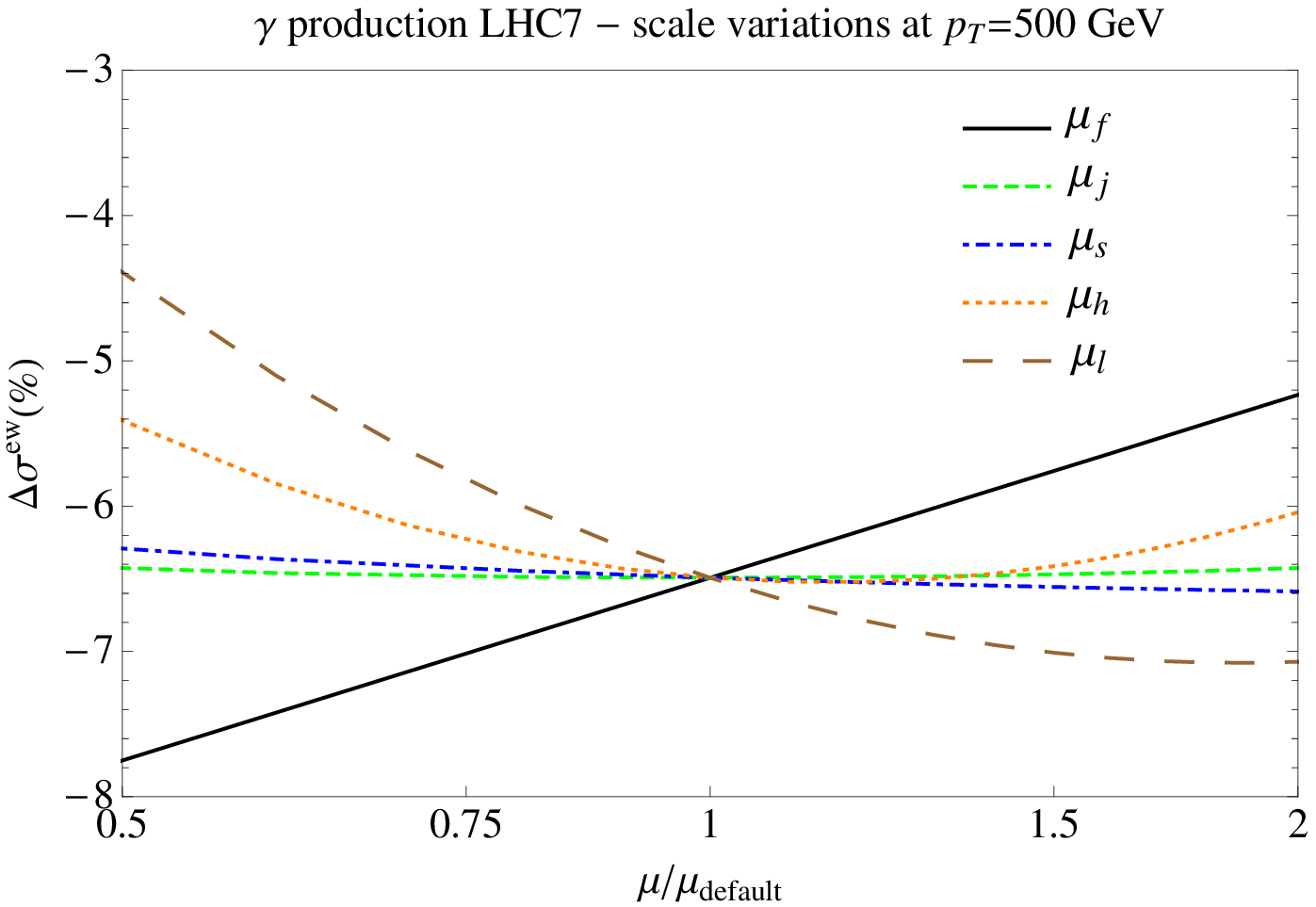}
\caption{Effect of scale variations in the cross section for $Z$ and $\gamma$ production at
  $p_T=500$~GeV as a reference point.}\label{fig:Zscalevars}
\end{figure}

To show the effect of including electroweak corrections to the cross
section, we plot in Fig.~\ref{fig:ZdiffewQCD} the difference
\begin{equation}\label{eq:diffsigew}
\Delta\sigma^{ew}\defeq\frac{\sigma_{ew}^{i}-\sigma^{i}}{\sigma^i},
\end{equation}
where $\sigma^i$ represents the
cross section with QCD corrections at order N$^i$LL, while $\sigma_{ew}^{i}$ also includes the electroweak corrections. The electroweak
corrections are always included at the same order, independently of the
value of $i$; i.e. including terms up to order $a$ in the exponent for
the hard function, and the leading corrections in the jet and soft
functions.
In the Figure, the black curves and bands correspond to $i=1$, and the green ones to
$i=2$. Again, in each plot we have varied the corresponding scale by a
factor of 2. The scales in the QCD and the electroweak parts are varied
simultaneously, both in the numerator and denominator of Eq.~(\ref{eq:diffsigew}). From the fact that the black and green curves are
almost identical, we learn that in order to study the
relative importance of the electroweak corrections it is not necessary to
include the QCD corrections at N$^2$LL accuracy. To better visualize
the effect of the scale variations, we choose a reference value $p_T=500\,{\rm GeV}$ and plot the cross section as a function of the deviation
from the default scale choices; this is shown in
Fig.~\ref{fig:Zscalevars}, for both the $Z$ boson and the photon.
 We observe that the electroweak corrections in the cross
section for photon production are smaller than those for $Z$ production. For our
final results, which are shown in Fig.~\ref{fig:Vallmu}, we add the bands
coming from the different scale variations in quadrature.

\section{\boldmath Results for $W$ production}\label{sec:W}
We now consider single $W$-boson production. The main difference with respect to $Z$ production is that
the $W^{\pm}$ bosons are charged. Therefore, both particles
in the final state, the $W^{\pm}$ and the quark or gluon, can have a jet function. This means that the factorization
formula is more complicated than in the previous section, and will
have the general form sketched in Eq.~(\ref{eq:schfact}).  The situation is similar to the study of QCD corrections to dijet
cross sections. A factorization formula for those processes was
derived in Ref.~\cite{Kidonakis:1998bk}, and the ingredients for N$^2$LL resummation in SCET were given in
Refs.~\cite{Kelley:2010fn,Kelley:2010qs}. The expression that we will use here has the same form as the one in Ref.~\cite{Kelley:2010qs}, but is
simpler, because we only have one possible group structure (the tree
level diagrams contain one ${SU(3)}_C$ and one $SU(2)_L$ matrix, but
those are in different spaces and commute), in contrast to the dijet case. Therefore, the $H$ and $S$ functions in the factorization
formula have no group indices.  We write the partonic cross section
for the channel $a\,b\to c\,W$ as
\begin{equation}
\hat{s}\frac{d\hat{\sigma}_{ab,W}}{d\hat{s}d\hat{t}}=\hat{\sigma}^B_{ab,W}H_{ab,W}(\hat{s},\hat{t},M_W,\mu)\int dp_W^2\,
dp_c^2\, dk\, J_W(p_W^2,\mu)J_c(p_c^2,\mu)S_{ab,W}(k,\mu)\delta\left(M_X^2+M_W^2-p_W^2-p_c^2-2E_Jk\right),
\end{equation}
and the hadronic cross section is then obtained after performing the convolution with the PDFs, analogous to Eq.~(\ref{eq:hadrcs}). Since we will only include electroweak
corrections to the jet and soft functions at leading order, $J_W$ reduces to a $\delta(p_W^2-M_W^2)$ times a prefactor, and we recover the structure for the factorization formula that we had in the
$Z$ case. In the following we give the results for the different ingredients of the factorization formula above.

The $W^{\pm}$ bosons in the broken theory come from the $SU(2)_L$ gauge bosons $W^{1,2}$
 in the unbroken theory, according to the
combination $W^{\pm}=1/\sqrt{2}(W^1\mp iW^2)$. Amplitudes with a $W^1$ or a
$W^2$ receive the same electroweak corrections as those
for $W^3$ in the previous section, i.e. the evolution factor
$\mathcal{U}_{ab}^{(W^3)}(\mu_h,\mu_l$) in Eq.~(\ref{eq:evHqq}). Below the scale $\mu_l$, the $W^{\pm}$
boson is treated as a field in HQET. The low-energy matching is given by\footnote{For simplicity, we use the subscript $q\bar{q}$ in $D_{q\bar{q}}^{(W^\pm\to W^\pm)}$, despite the fact that the two quarks have different flavor.}
\begin{equation}\label{eq:defDW}
D_{q\bar{q}}^{(W^\pm\to W^\pm)}(\mu)=D_{q\bar{q}}^{(W^3\to
  Z)}(\mu)+\frac{\alpha_2}{4\pi}\log\frac{\hat{s}}{\mu^2}\left[-
\log\frac{M_W^2}{\mu^2}+c_W^2 \log\frac{M_Z^2}{\mu^2}\right],
\end{equation}
and the running in the low-energy broken theory is given by the corresponding factor, $\mathcal{U}_{q\bar{q}}^{(W^{\pm})}\left(\mu_l,\mu_f\right)$, with the
following anomalous dimensions
\begin{eqnarray}
\Gamma^{(W^{\pm})} & = &
\frac{\alpha_{em}}{4\pi}\frac{\Gamma_0}{2}\left(Q^2+Q'^2\right),\\
\gamma_{q\bar{q}}^{(W^{\pm})} & = & \frac{\alpha_{em}}{4\pi}\left[-\Gamma_0\frac{1}{2}\log\frac{M_W^2}{\hat{s}}\right.\left.-\Gamma_0 QQ'i\pi+\Gamma_0 QQ_{W^{\pm}}\log\frac{-\hat{t}}{\hat{s}}-\Gamma_0 Q'Q_{W^{\pm}}\log\frac{-\hat{u}}{\hat{s}}-3\left(Q^2+Q'^2\right)-2\right],
\end{eqnarray}
where $Q$ is
the charge of the quark and $-Q'$ the charge of the antiquark
(i.e. $Q-Q'=Q_{W^{\pm}}$, with $Q_{W^{\pm}}$ the charge of the
$W^{\pm}$ boson). Since the $W^\pm$ is massive in the low-energy theory, the expression Eq.~(\ref{magic}) cannot be used to obtain the above anomalous dimension. The appropriate expression for the massive case was given in \cite{Becher:2009kw}. The hard function times Born-level cross section is then given by
\[
\hat{\sigma}^B_{q\bar{q},W^{\pm}}H_{q\bar{q},W^{\pm}}(\hat{s},\hat{t},M_W,\mu_f)=\frac{\pi\alpha_s(\mu_h)}{\hat{s}}\frac{2C_F}{N_c}\left(\frac{\hat{t}^2+\hat{u}^2+2M_W^2\hat{s}}{\hat{t}\hat{u}}\right)\left|\mathcal{U}_{q\bar{q}}^{(W^{\pm})}\left(\mu_l,\mu_f\right)\right|^2
\]
\begin{equation}\label{eq:evHWqq}
\times\left|\frac{V_{i j}}{\sqrt{2}}\sqrt{\alpha_2(\mu_h)} \mathcal{D}_{q\bar{q}}^{(W^\pm\to W^\pm)}\left(\mu_l\right)\mathcal{U}_{q\bar{q}}^{(W^\pm)}\left(\mu_h,\mu_l\right)\right|^2 H_{q\bar{q},W^{\pm}}(\hat{s},\hat{t},M_W,\mu_h),
\end{equation}
Here, $V_{i j}$ denotes the Cabibbo-Kobayashi-Maskawa (CKM) matrix and we have assumed that the quark and anti-quark are from generations $i$ and $j$, respectively.
%
Like in the $Z$ case
of the previous section, we do not write pure-QCD corrections
explicitly, and $H_{ab,W}(\hat{s},\hat{t},M_W,\mu_h)$ is again 1 plus
pure-QCD terms at the order we are working. The expression for the Compton channel can be obtained from the result in Eq.~(\ref{eq:evHWqq}) above by crossing.

The $W^{\pm}$ jet function is defined in HQET, see Ref.~\cite{Fleming:2007xt}. At the order we need here we have
\begin{equation}\label{eq:jetW}
J_W(p^2,\mu_f)=e^{-4S(\mu_j,\mu_f)}\delta\left(p^2-M_W^2\right),
\end{equation}
with $S(\nu,\mu)$ given by Eq.~(\ref{eq:defSA}) with
$\Gamma=\frac{\alpha_{em}}{2\pi}$. Finally, for the
leading electroweak corrections to the soft functions
that we need in this case we obtain
\begin{equation}\label{eq:softW}
S_{q\bar{q},W}(k,\mu_f)=e^{-4\left(Q^2+Q'^2-1\right)S(\mu_s,\mu_f)}\delta (k)\quad;\quad S_{qg,W}(k,\mu_f)=e^{-4\left(Q^2-Q'^2-1\right)S(\mu_s,\mu_f)}\delta(k),
\end{equation}
again with $S(\nu,\mu)$ given by Eq.~(\ref{eq:defSA}) with
$\Gamma=\frac{\alpha_{em}}{2\pi}$. 

\begin{figure}
\centering
\includegraphics[width=8cm]{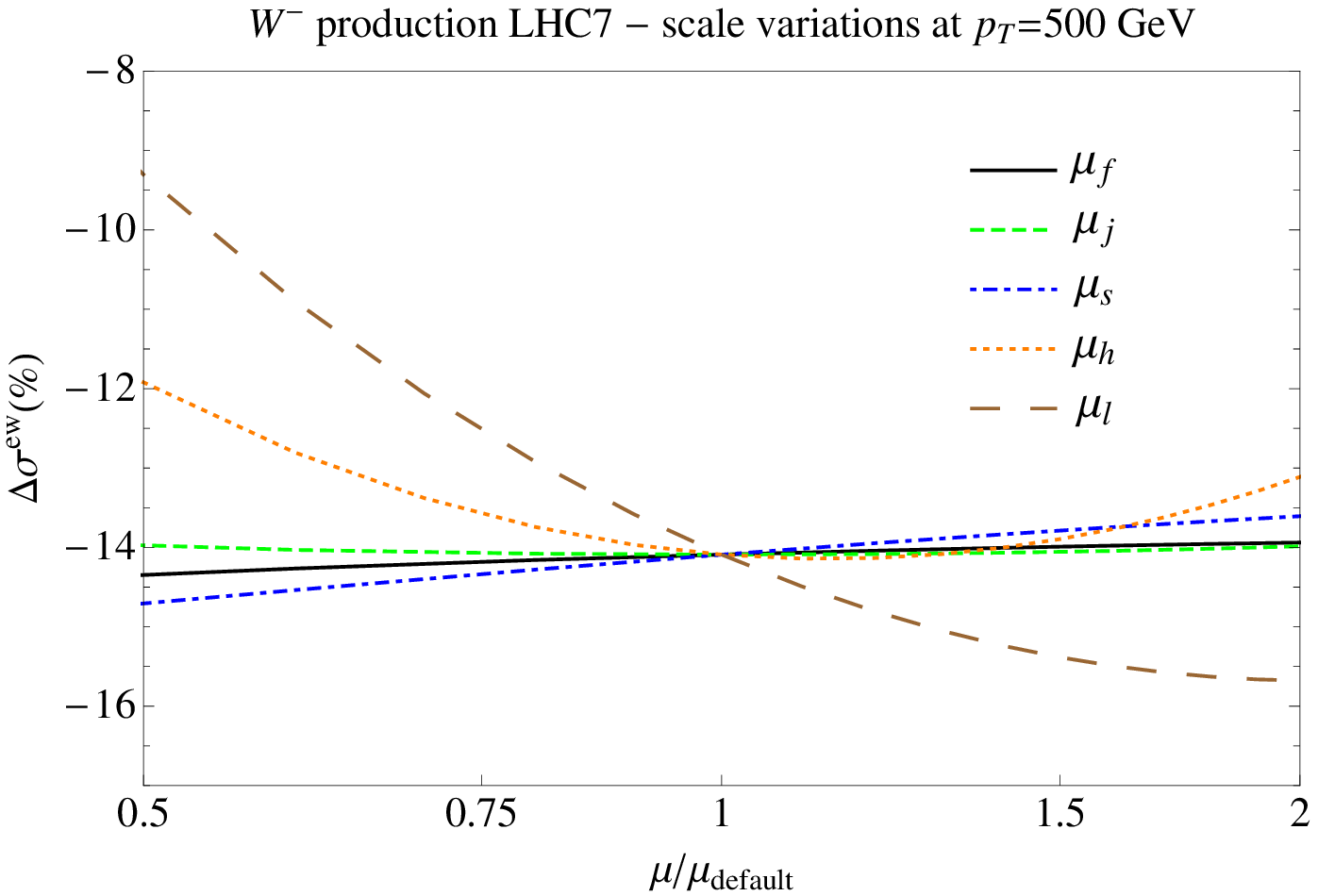}\phantom{ab}
\includegraphics[width=8cm]{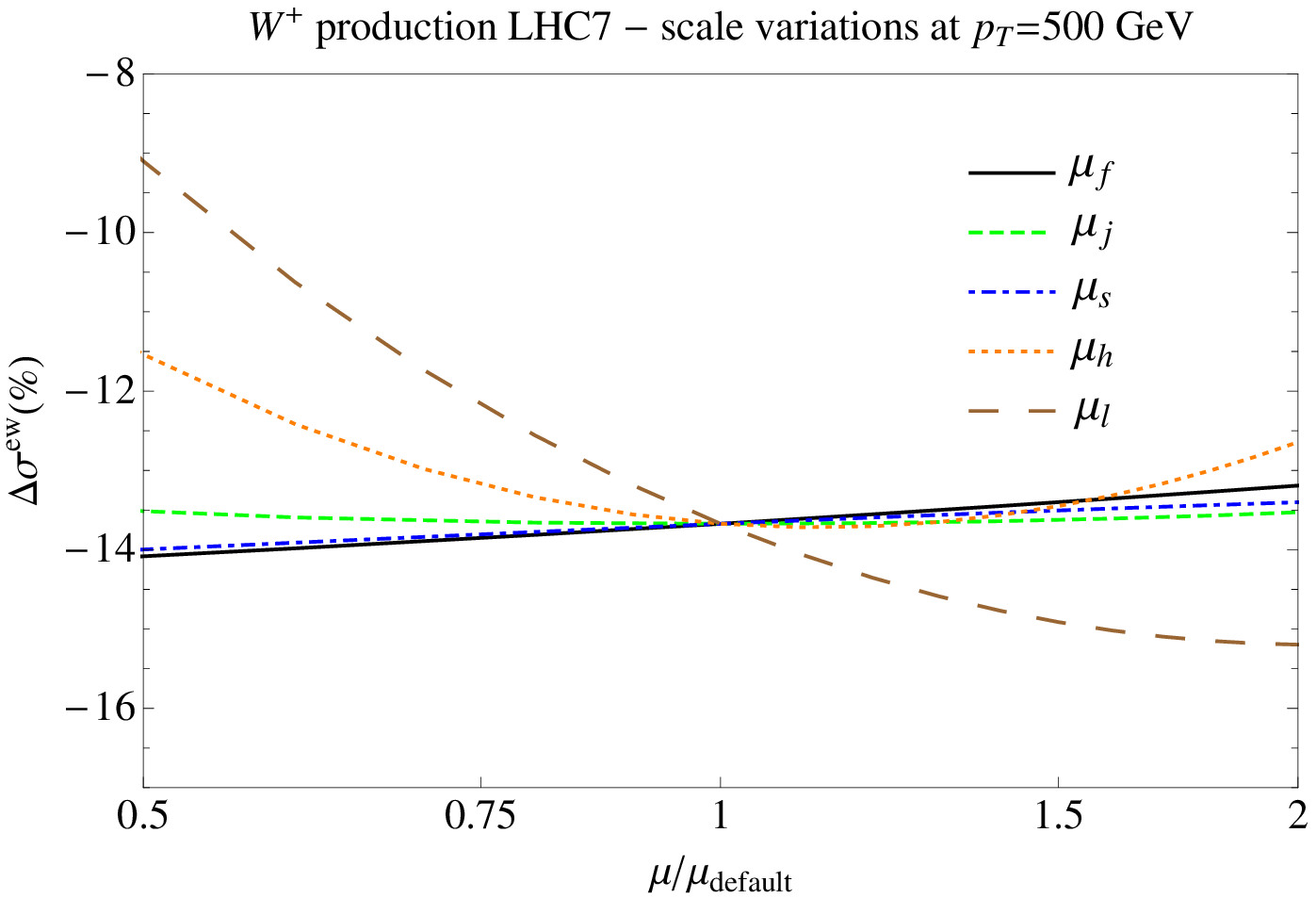}
\caption{Effect of scale variations in the cross section for $W^-$ and
  $W^+$ production at
  $p_T=500$~GeV as a reference point.}\label{fig:Wscalevars}
\end{figure}

We show the effect of the different scale variations for $W^{\pm}$
production in Fig.~\ref{fig:Wscalevars}. The final results for the
cross section are shown in Fig.~\ref{fig:Vallmu}. The results for $W^\pm$ are numerically quite similar to the $Z$-boson case.

\section{Summary and conclusions}\label{sec:disc}

\begin{figure}
\centering
\renewcommand{\arraystretch}{2.5}
\begin{tabular}{lr}
\includegraphics[width=8cm]{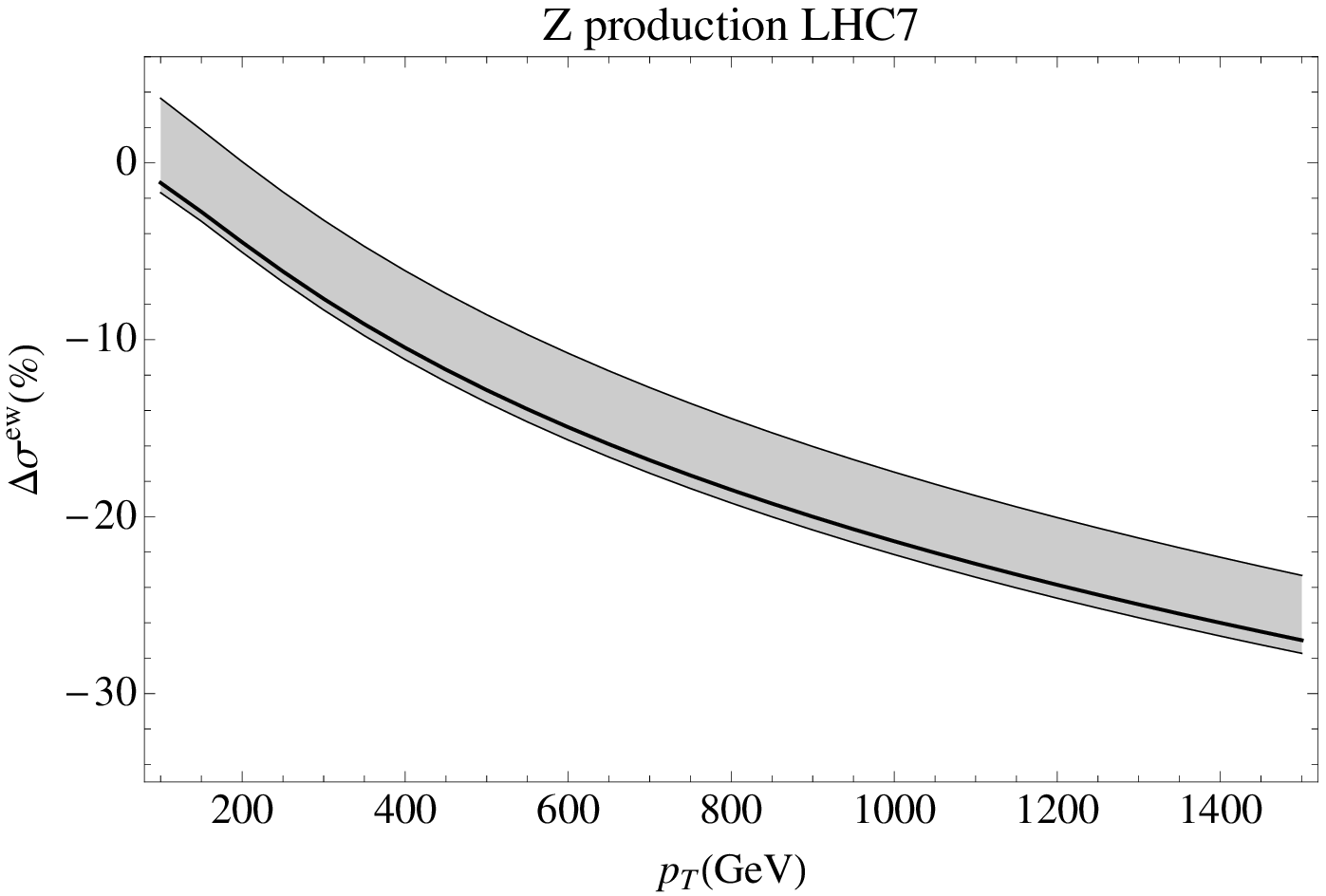}\phantom{ab} &
\includegraphics[width=8cm]{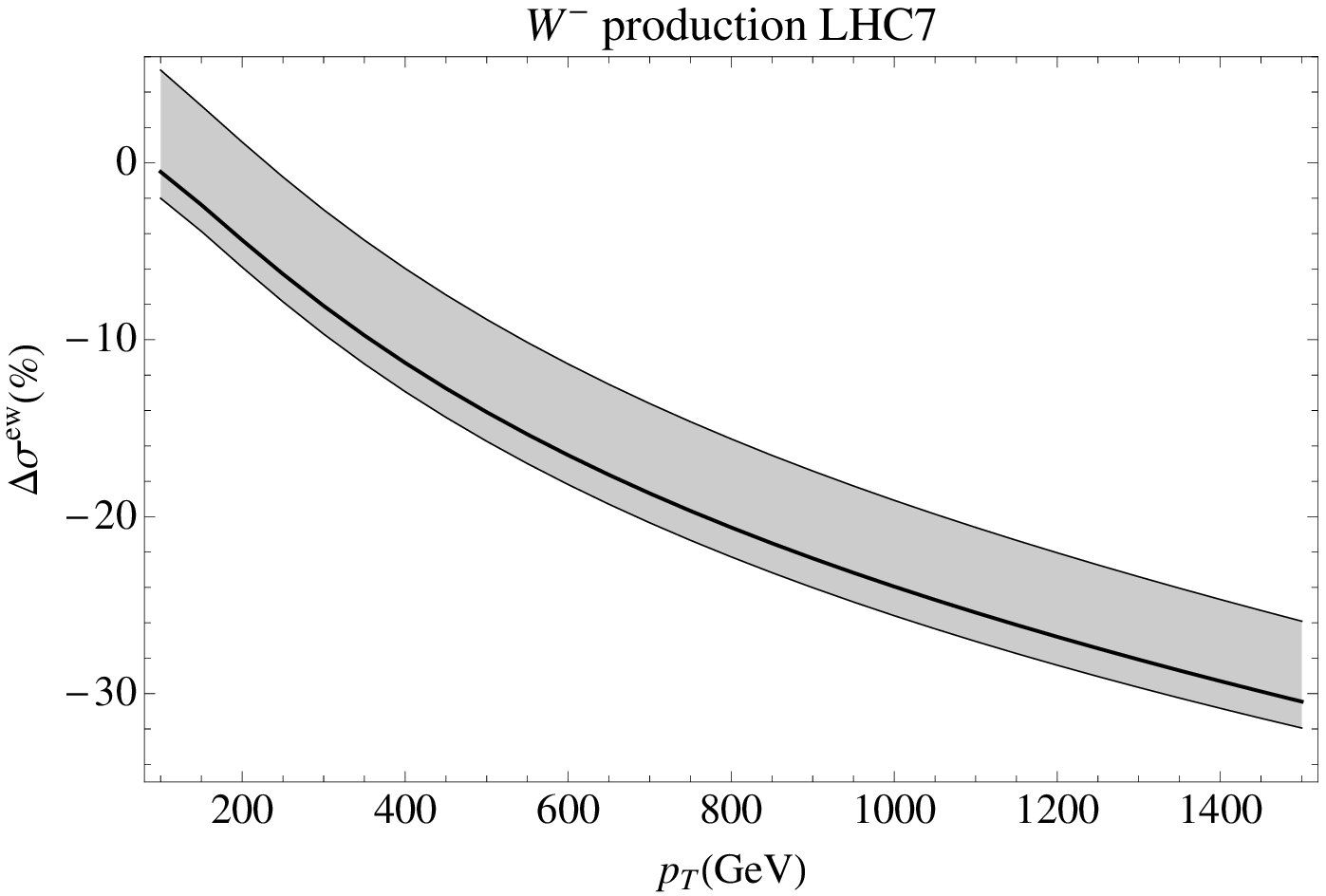}\\
\includegraphics[width=8cm]{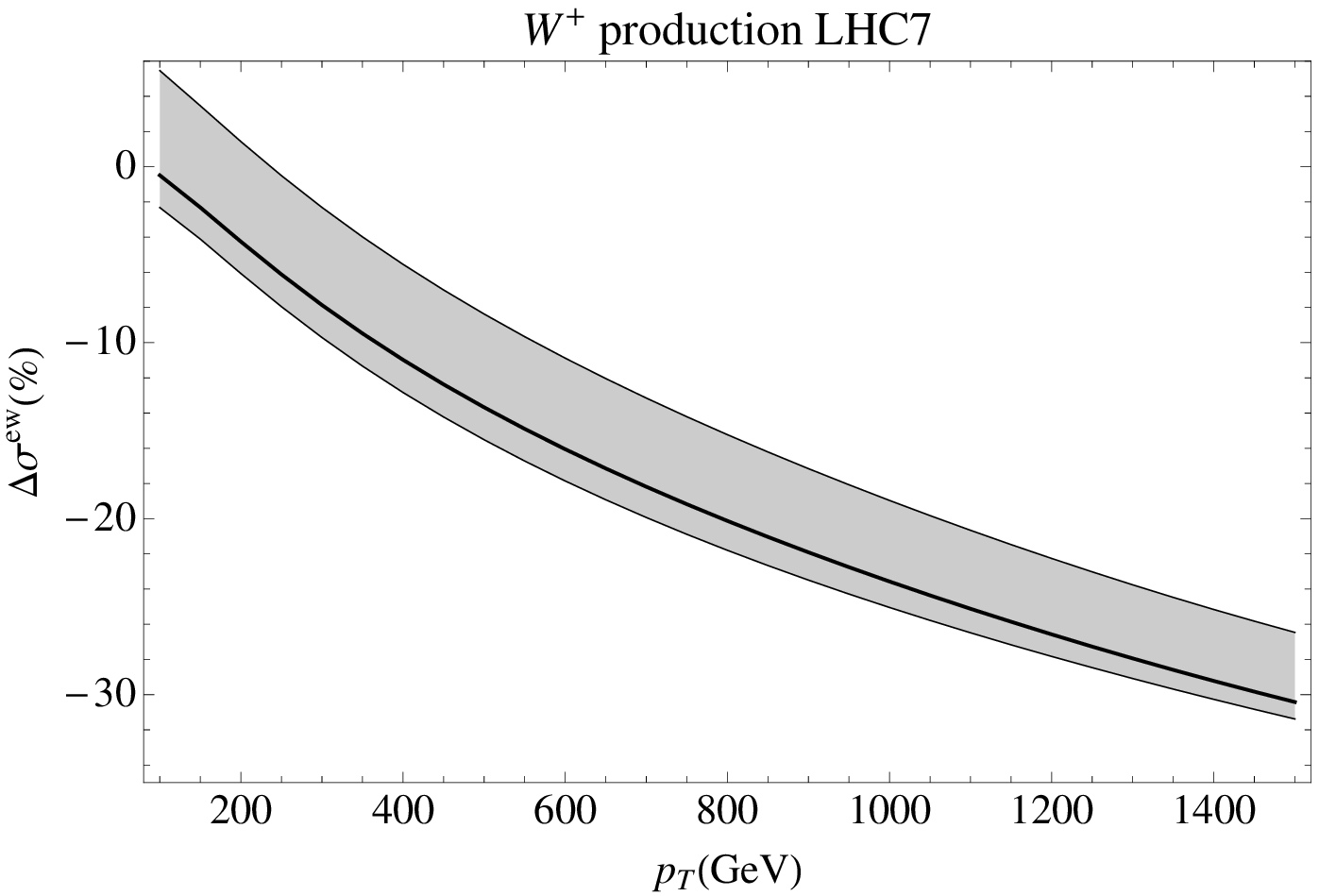}\phantom{ab} &
\includegraphics[width=8cm]{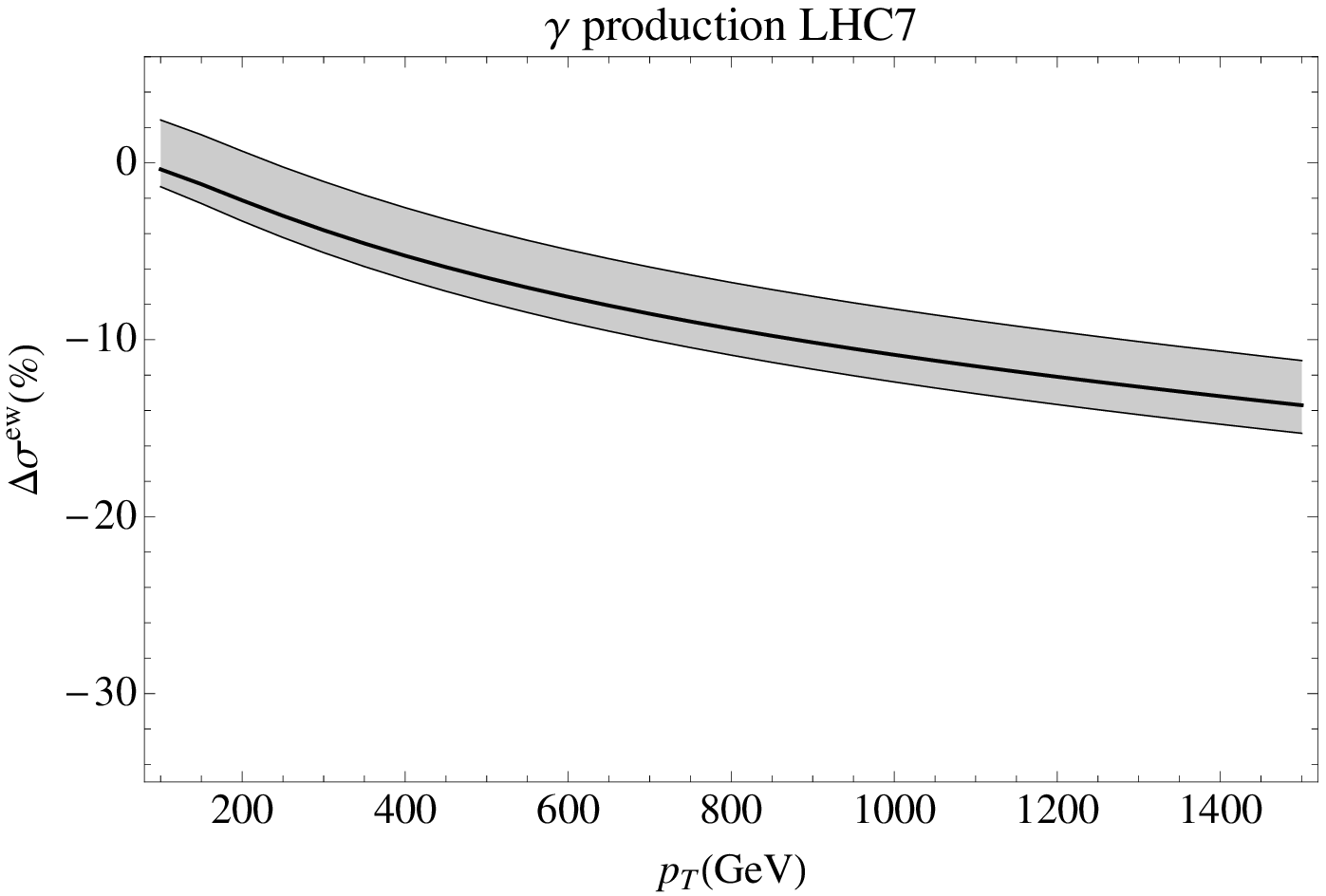}
\end{tabular}
\caption{$W$, $Z$ and $\gamma$ production for the LHC 7 TeV. We plot the difference of
  cross sections with and without electroweak corrections given by
  Eq.~(\ref{eq:diffsigew}), with $i=1$. The bands reflect the
  perturbative uncertainty of the results. They are obtained by
  first varying each of the scales appearing in the factorization formula by
a factor of 2 (as discussed in the text), and then adding these different individual bands in quadrature.}\label{fig:Vallmu}
\end{figure}

In summary, in this paper we have computed electroweak Sudakov
corrections to the cross sections for single $W$, $Z$ and $\gamma$ production
at large transverse momentum, within SCET using the approach of
Refs.~\cite{Chiu:2009mg,Chiu:2009ft}. We
have presented complete results for the factorized hadronic cross sections with electroweak
corrections. At the LHC, these corrections are of order $20\%$ for $p_T\sim1$~TeV for $Z$ and $W$ production and about half as big for prompt photon production. Their inclusion is necessary to obtain precise predictions of the spectrum at large $p_T$. In our numerical analysis, we included both QCD and electroweak corrections, and discussed the most adequate way to set the different scales that
appear in the factorized form for the cross section. Our results are
summarized in Fig.~\ref{fig:Vallmu}, where we show the effect of
including electroweak corrections in the cross section by plotting the
difference of cross sections with and without electroweak corrections. 

Two important features of our results are the following: first of all, the main part
of the electroweak corrections is contained in the hard function, and
the effects on the jet and soft functions are much smaller. This is 
also evident from the fact that the bands due to $\mu_j$ and $\mu_s$
variation in Fig.~\ref{fig:ZdiffewQCD} are much smaller
than the ones coming from the variation of the other scales. This result
is in accordance with the statements made in Ref.~\cite{Kuhn:2007cv}
regarding the small impact of the different treatments of singularities
in real radiation photon diagrams on the size of the corrections \cite{Hollik:2007sq,Kuhn:2007cv}. 
The second feature worth stressing is that the relative importance of the
electroweak corrections, as defined in Eq.~(\ref{eq:diffsigew}), does not
depend much on the order to which we work in the QCD part. This means that
one can, to good accuracy, include electroweak effects via an overall prefactor in existing pure-QCD computations.

Electroweak Sudakov corrections to vector-boson production have been considered before. In particular, Refs.~\cite{Kuhn:2004em,Kuhn:2005gv,Kuhn:2007cv} have presented analytic
expressions for the IR-finite part of the virtual electroweak
corrections, at next-to-leading logarithmic accuracy up to two loops, in the limit $M_{W,Z}^2 \ll \hat{s} \sim\hat{t}\sim\hat{u}$, for $Z$, $\gamma$ and $W^{\pm}$
production, respectively. These terms correspond to our expressions
for $\hat{\sigma}^B_{q\bar{q},V}H_{q\bar{q},V}$ expanded to
order $\alpha_i^3$, with
$\mu_l=\mu_f=M_{W}=M_Z$. Refs.~\cite{Kuhn:2004em,Kuhn:2005gv,Kuhn:2007cv} do not  consider QCD corrections and the mixing terms $S_{\alpha_s\alpha_i}(\nu,\mu)$ are therefore not included in their results. Switching off the QCD terms in our result and performing a fixed-order expansion, we find agreement with their results.

The results of this paper together with the resummation of the pure-QCD corrections,
which can be performed at N$^3$LL accuracy, yield predictions for single electroweak boson
production at large transverse momentum at an unprecedented level of
accuracy. Ratios of these $p_T$ spectra can be used to constrain the
$u/d$ ratio of PDFs or as a theoretical input in estimations of the
$Z(\to \nu\bar{\nu})+$jets background to new physics searches, as
recently discussed in Ref.~\cite{Malik:2013kba}. A comprehensive study comparing with available LHC data, including N$^3$LL accuracy for the pure-QCD resummation, will be the subject of a future publication.

\begin{acknowledgments}
This work is supported by the Swiss National Science Foundation (SNF)
under grant 200020-140978 and the Sinergia grant number
CRSII2\underline{ }141847\underline{ }1.

\end{acknowledgments}

\appendix

\section{Beta functions in the SM}\label{sec:appbSM}
The two-loop running of a general direct-product group can be found in Ref.~\cite{Jones:1981we}.
Recently, the running of the couplings in the SM up to three loops has
also been computed, in Refs.~\cite{Mihaila:2012fm,Mihaila:2012pz,Bednyakov:2012rb}. For convenience this Appendix collects the expressions for the SM beta functions
that are used throughout the paper. 

We write the beta function for
the coupling $\alpha_a$ as
\begin{equation}\label{eq:defbfun}
\beta(\alpha_a)=-2\alpha_a\left[\beta_0^{\alpha_a}\frac{\alpha_a}{4\pi}+\beta_1^{\alpha_a}\left(\frac{\alpha_a}{4\pi}\right)^2+\beta_1^{\alpha_a\alpha_b}\frac{\alpha_a}{4\pi}\frac{\alpha_b}{4\pi}+\cdots\right].
\end{equation}
The coefficients that are used in the paper read
\begin{eqnarray}
\beta_0^{\alpha_1} & = &
\frac{5}{3}\left(-\frac{4}{3}n_g-\frac{1}{10}n_h\right)=-\frac{41}{6}, \nonumber\\
\beta_0^{\alpha_2} & = &
\frac{22}{3}-\frac{4}{3}n_g-\frac{1}{6}n_h=\frac{19}{6},\nonumber\\
\beta_1^{\alpha_s\alpha_1} & = & -\frac{5}{3}\frac{11n_g}{30}=-\frac{11}{6}\, ,\nonumber\\
\beta_1^{\alpha_s\alpha_2} & = & -\frac{3n_g}{2}=-\frac{9}{2}\, ,\\
\beta_1^{\alpha_1\alpha_s} & = & -\frac{5}{3}\frac{44n_g}{15}=-\frac{44}{3}\, ,\nonumber\\
\beta_1^{\alpha_2\alpha_s} & = & -4n_g=-12\, ,\nonumber\\
\beta_0^{\alpha_{em}} & = &
-\frac{4}{3}\left[ N_c\left(3Q_d^2+2Q_u^2\right)+3Q_l^2 \right]=-\frac{80}{9}\,,\nonumber\\
\beta_1^{\alpha_s\alpha_{em}} & = &
2\left(3Q_d^2+2Q_u^2\right)=-\frac{22}{9}\,, \nonumber\\
\beta_1^{\alpha_{em}\alpha_s} & = & -4C_F\left[ N_c\left(3Q_d^2+2Q_u^2\right)\right]=-\frac{176}{9}\, , \nonumber
\end{eqnarray}
where $n_g=3$ is the number of generations, $n_h=1$ is the number of Higgs
doublets, $C_F=(N_c^2-1)/(2N_c)$, $N_c=3$ is the number of
colors, $Q_d=-1/3$ and $Q_u=2/3$ are the charges of the
down- and up-type quarks respectively, and $Q_l=-1$ is the charge
of the charged leptons.

\end{document}